\documentclass[onecolumn]{aa}
\usepackage{graphicx}
\usepackage{amsmath}
\usepackage[normalem]{ulem}
\newcommand{\ISO}{{\em ISO }}
\newcommand{\ISOc}{{\em ISO}}
\newcommand{\ISOCAM}{{\em ISOCAM }}
\newcommand{\ISOCAMc}{{\em ISOCAM}}
\newcommand{\Spitzer}{{\em Spitzer }}
\newcommand{\Spitzerc}{{\em Spitzer}}
\newcommand{\Herschel}{{\em Herschel }}

\begin{document}
  \title{The 12 $\mu$m \ISOc-ESO-Sculptor 
   \thanks{Based on observations with \ISO, an ESA project with instruments 
      funded by ESA member states (specially the PI countries France, 
      Germany, The Netherlands and the United Kingdom) with the participation 
      of ISAS and NASA. The ESO-Sculptor Survey is based on observations 
      collected at the European Southern Observatory (ESO), La Silla, Chile.}
  and 24 $\mu$m \Spitzer faint counts reveal a population of ULIRG/AGN/dusty 
massive ellipticals}
  
\subtitle{Evolution by types and cosmic star formation}
  
\titlerunning{$ISO$-ESS Counts at 12$\,\mu$m}  
\authorrunning{Rocca-Volmerange, Lapparent, Seymour \& Fioc}
 
\author{B. Rocca-Volmerange\inst{1,2}, V. de Lapparent\inst{1}, 
N. Seymour\inst{1,3} \and M. Fioc \inst{1}
    }
\offprints{B. Rocca-Volmerange} 
\institute{Institut d'Astrophysique de Paris, UMR7095 CNRS, Univ. Pierre \& Marie Curie, 98 bis boulevard Arago, 75014 Paris, France\\
   \email{brigitte.rocca@iap.fr}
\and 
   Universit\'e Paris Sud, B\^at 121, F-91405 Orsay Cedex, France\\
\and
Spitzer Science Center, California Institute of Technology,
    Mail Code 220-6, 1200 East California Boulevard, Pasadena, CA 91125 USA.}
  
\date{submitted: February, 2006}
  
  %%%%%%%%%%%%%%%%%%%%%%%%%%%%%%%%%%%%%%%%%%%%%%%%%%%%%%%%%%%      

 \abstract{Multi-wavelength galaxy number  counts provide clues on the
nature of galaxy evolution. The  interpretation per galaxy type of the
mid-IR faint counts obtained  with \ISO and \Spitzerc, consistent with
the  analysis of  deep  UV-optical-nearIR source  counts, provide  new
constraints on the dust  and stellar emission.  Discovering the nature
of  new  populations,  as ultra-luminous  ($\geq$10$^{12}$L$_{\odot}$)
infrared galaxies  (ULIRGs), is also crucial  for understanding galaxy
evolution  at high  redshifts.}  {We  first present  the  faint galaxy
counts  at 12$\,\mu$m  from  the catalogue  of the  \ISOc-ESO-Sculptor
Survey  (\ISO-ESS) published in  a companion  article (Seymour  et al.
2007), which go down to 0.24 mJy after corrections for incompleteness.
We  verify the  consistency with  the existing  \ISO number  counts at
15$\,\mu$m. Then we analyse them as well as the 24$\,\mu$m ($Spitzer$)
faint  counts to  constrain  the  nature of  ULIRGs,  the cosmic  star
formation history  and the time-scales  for mass build-up.}   {We show
that the ``normal'' scenarios in our evolutionary code P\'EGASE, which
had   previously  fitted  the   deep  UV-optical-nearIR   counts,  are
unsuccessful  at 12$\,\mu$m  and  24$\,\mu$m. We  then  propose a  new
scenario of ULIRG adjusted to the observed cumulative and differential
12$\,\mu$m and 24$\,\mu$m counts  and based on observed 12$\,\mu$m and
25$\,\mu$m IRAS luminosity functions and mid-IR colors from P\'EGASE.}
{We succeed in modelling the typical excess simultaneously observed at
12$\,\mu$m,  15$\,\mu$m  (\ISOc)  and  24$\,\mu$m (\Spitzerc)  in  the
cumulative  and differential  counts by  only changing  9\%  of normal
galaxies  (1/3 of  the ellipticals)  into ultra-bright  dusty galaxies
evolving according to the  scenario of ellipticals, and interpreted as
ULIRGs.   These objects  present similarities  with the  population of
radio-galaxy  hosts  at  high  $z$.  No number  density  evolution  is
included  in our  models even  if  rare occasional  starbursts due  to
galaxy  interactions  remain compatible  with  our results.}   {Higher
spectral  and   spatial  resolution   in  the  mid-IR   together  with
submmillimeter  observations using  the  future \Herschel  observatory
will  be  useful  to  confirm  these  results.}   \keywords{Cosmology:
surveys;  Galaxies:  evolution -  spiral  -  infrared  - photometry  }
\maketitle

  %%%%%%%%%%%%%%%%%%%%%%%%%%%%%%%%%%%%%%%%%%%%%%%%%%%%%%%%%%%

  \section{Introduction} 

  The mid-infrared  extra-galactic source counts provide  clues on the
  evolution of galaxies at high $z$  and allow us to follow the cosmic
  star formation history up to  when the Universe was only one quarter
  of its present age.
% up to the origin of galaxies. 
  The  high sensitivity  of the  infrared deep  surveys  observed with
  ESA's {\em  Infrared Space Observatory ISO} (Kessler  et al.  2003),
  and  more recently  with \emph{Spitzer  Space Telescope}  (Werner et
  al. 2004)  offers an unique  opportunity to study the  obscured star
  formation  process through the  emission of  grains heated  by young
  stars and possibly  by an active nucleus, thus  reaching galaxies at
  their  earliest  epochs.   Interactions  of galaxies  are  known  to
  explain the  huge IR emission  detected in Ultra Luminous  Infra Red
  Galaxies (ULIRG; Soifer et al.  1984).  One crucial issue is whether
  the ULIRGs  provide observational evidence  that galaxy interactions
  play a fundamental role  in galaxy evolution.  Moreover, because the
  infrared  luminosity  depends  on  the dust  mass  accumulated  from
  stellar ejecta,  it provides complementary diagnostics  of past star
  formation  which may in  turn be  used to  model the  star formation
  history. Over the years, a long  series of deep surveys from the UV
  to the optical,  down to the extreme depth of  the Hubble Deep Field
  North (HDF-N) at $B=29$ (Williams et al. 1996), have constrained the
  direct stellar emission in terms  of cosmology and scenarios of star
  formation.   Fioc  and Rocca-Volmerange  (1999a)  derived  a set  of
  galaxy  populations  fitting  the multi-wavelength  (UV-optical-near
  infrared) deep galaxy counts dominated by stellar emission. This set
  defines the  evolution scenarios of  eight various galaxy  types and
  their  number  fractions.  In  the  mid-infrared,  galaxy  light  is
  dominated by the dust emission  from grains in the form of graphite,
  silicates or Polycyclic Aromatic Hydrocarbons (PAH; Puget \& L\'eger
  1989).   Because different  time-scales characterise  the respective
  star  and dust emissions,  the results  of optical  and mid-infrared
  source counts may significantly differ.

  However two difficulties hamper  the interpretation of data.  One is
  the  lack  of  homogeneity  between  the  various  surveys:  due  to
  large-scale  galaxy clustering,  the statistical  properties derived
  from  deep  pencil  beam  surveys suffer  from  ``cosmic  variance''
  (Somerville et al. 2004), and might thus differ from the analyses on
  large area  surveys.  Another difficulty  is the variety  of sources
  (starbursts due to mergers, normal evolved galaxies, AGNs) and their
  intrinsic  evolution. As an  example, the  time-scale  of star
  formation associated to galaxy interactions is significantly shorter
  ($\simeq$10$^8$yrs)   than  the  star   formation  time-scales   of  galaxy
  populations  ($>10^9$yrs, depending  on spectral  type)  observed in
  deep surveys. We consider the contribution of AGNs to be minimal at
  the flux density range explored  here, but the impact of an embedded
  hidden AGN is discussed below.

Our  large  area  \ISOCAM  ESO-Sculptor  Survey  (\ISOc-ESS)  at
12$\,\mu$m, published  in the companion paper (Seymour  et al, 2007),
here tackles these various limitations by covering a significant area,
and  by  using  the new  code  P\'EGASE.3  (Fioc  et al.  2007)  which
coherently  predicts the  evolving  stellar and  grain emissions  from
evolved galaxies as  well as young starbursts.  It  is able to predict
\emph{both}  starlight  and dust  emission  from  the  optical to  the
thermal  infrared,  by  taking  into  account  the  transfer  and  the
reprocessing  of light in  the different  wavelength domains. A
large  variety of  star  formation time-scales  is  considered in  our
evolutionary  templates but no evolution  of the  number  density of
galaxies is included in the model.

  Several  other major surveys in the mid-infrared have been 
  performed with  the \ISOCAM
  camera which also provide deep  galaxy counts. The largest survey is
  ELAIS (Rowan-Robinson et  al. 1999, 2004), which covers  12 sq. deg.
  in the flux  range 0.45--150 mJy at wavelengths  of 6.7$\,\mu$m (LW2)
  and 15$\,\mu$m (LW3); the corresponding galaxy counts were published
  by Serjeant  et al.  (2000).   Following the preliminary  surveys of
  Taniguchi et  al.  (1997) and  Oliver et al.  (1997),  several major
  surveys have  also been performed which provide  deep galaxy counts.
  Among them are the \ISO  15$\,\mu$m observations of the Lockman Deep
  Field and  the Marano-ROSAT Ultra  Deep Field (Aussel et  al.  1999,
  Elbaz  et  al.  1999)  and  the Lockman  Shallow  Field  (Flores  et
  al. 1999). Even deeper surveys have been obtained in the LW2 and LW3
  filters, centered on  Abell cluster 2390 (Altieri et  al. 1999), the
  HDF field, and others fields  covering areas of 2.5 arcmin in radius
  (Oliver et al.   2000, 2002), and of 16 arcmin$^2$  (Sato et al. 2003).
  These  various  surveys  yield  galaxy number counts  which  are  in
  reasonable agreement, and detect an excess in the number of galaxies
  at faint fluxes (below $\sim1$ mJy) which is often interpreted as an
  increase  in  the  star  formation  rate with  look-back  time.  For
  example, Pozzi et al.  (2004) fit the 15$\,\mu$m number counts
  by  introducing a population  of evolving  starbursts, based  on the
  local  star-forming  prototype  M82  (Silva  et  al.  1998),  which
  undergoes very strong  luminosity or density evolution parameterized
  as $(1+z)^\alpha$, with $\alpha\simeq$3--4.

All the differential 24$\,\mu$m counts also show a systematic
departure from an Euclidean Universe at fluxes fainter than $\sim$\ 0.5
mJy (Rodighiero et al. 2006), with a peak at $\sim$\ 0.3 mJy, similarly
to the 15$\,\mu$m and 12$\,\mu$m 
%(Fig.~\ref{figure:Diff_nomodel})
observations. Below 60$\,\mu$Jy, the 24$\,\mu$m galaxy number counts
obtained by the \Spitzerc/MIPS deep surveys (Marleau et al. 2004,
Papovich et al. 2004, Chary et al. 2005) face the problem of confusion
limit due to extragalactic sources. Several interpretations propose
huge evolution factors in luminosity and/or density. In the
\emph{Chandra} Deep Field South survey, the sample of 2600
\Spitzerc/MIPS sources brighter than ~80$\mu$Jy is analyzed
with the comoving IR energy varying as $(1+z)^{3.9\pm 0.4}$ (Le Floc'h
et al. 2005).  The origin of such a high evolution is however not
described.  More recently, Caputi et al. (2006) analyzed the stellar
populations of the \Spitzerc/MIPS 24$\,\mu$m galaxies in the
GOODS/CDFS from $K_s$ images. They show evidence for a bump in the
redshift distribution at $z=1.9$, induced by a significant population
of galaxies with PAH emission.

  The  combined \ISOc-ESS presented here
  has comparable depth ($\sim80$\%  completeness at $\sim0.7$ mJy) and
  surface  area (680  arcmin$^2$)  at 12$\,\mu$m  as the  intermediate
  surveys  (Lockman   Deep  and  Marano  Deep   fields)  performed  at
  $15\,\mu$m  as part  of  the \ISOCAM  Guaranteed Time  Extragalactic
  Surveys.  The specificity of the \ISOc-ESS survey is  to provide deep
  galaxy counts at $\sim12\pm3.5\,\mu$m (LW10 filter), in a wavelength
  range where  the spectral energy distribution (SED)  is dominated by
  the  signatures of  PAH emission;  it thus  strongly  constrains the
  evolution  process.   The  target   field  is  located   within  the
  ESO-Sculptor Survey (de Lapparent et al.  2003, 2004), for which deep
  optical  $BVR_\mathrm{c}$ magnitudes  up to  $R_\mathrm{c}\le23.5$\ and
 spectroscopic  redshifts  at  $R_\mathrm{c}\le21.5$  have  been obtained
  (Arnouts et al. 1997, Bellanger et al.  1995a).
   
  We then  model the  12$\,\mu$m number counts  using the  new version
  P\'EGASE.3  of  the ``Projet  d'Etude  des  GAlaxies par  Synth\`ese
  Evolutive''(Fioc et al.  2007; Fioc \& Rocca-Volmerange 1997, 1999b;
  see  also  www2.iap.fr/pegase),  which  coherently  complements  the
  UV-optical-NIR  emission  from  stars  and  gas with  the  mid-  and
  far-infrared emission  from dust. The specific goal  of our analysis
  is to predict the mid-infrared number counts using firstly the
  same scenarios  of galaxy  evolution by type,  with the  same number
  fractions and  the same total  number densities of galaxies  as used
  for the successful predictions  of the UV-optical-NIR deep counts by
  Fioc \&  Rocca-Volmerange (1999a). Secondly, if  needed, other
  scenarios are  proposed to  model the
  population  of ULIRGs  able  to reproduce  the systematic  departure
  observed around ~0.3mJy in mid-infrared surveys.

Section  2  describes  the  parameters  of the  \ISOc-ESS  field  
  observed with the large pass-band \ISOCAM LW10/12$\,\mu$m filter and
  summarises  the data  analysis required  to extract  a  catalogue of
  sources presented  in a companion article (Seymour  et al.  2007). The
  corresponding   cumulative  and   differential  galaxy   counts at 
12$\,\mu$m  are
  presented in  section 3. We resume in section 4  the parameters of
  two models  of evolution scenarios and the  observed IRAS luminosity
  function  at 12$\,\mu$m,  tentatively used  to model  galaxy counts.
  The respective fits of the cumulative and differential counts of the
  ESO-ESS survey  by the two models  are compared in  section 5. 
  The best model 2 identifies a
  population  of  ultra  bright  ellipticals  interpreted  as  distant
  ULIRGs.  Applied   to  24$\,\mu$m   number  counts  from   the  deep
  \Spitzerc/MIPS  surveys,  we  show   in  Section  6  that  the  same
  population  of ultra  bright ellipticals also reproduces  the 24$\,\mu$m
  differential excess, making robust our model 2.  Section 7 presents the
  cosmic  star formation  history resulting  from the  best  fit, taking 
  into account the respective 
  contributions per galaxy type.   Section 8 discusses the stellar and
  dust masses of the revealed ULIRG population, and the possibility of
  an hidden AGN. The final section presents our conclusions.

  %%%%%%%%%%%%%%%%%%%%%%%%%%%%%%%%%%%%%%%%%%%%%%%%%%%%%%%%%%%

  \section{Observations and data reduction}

  %==========================================================
  \subsection {The ESO-Sculptor Survey field}

  The selected field, the ESO-Sculptor Survey (ESS) of faint galaxies,
  is  described  in  its  complete  version by  de  Lapparent  et  al.
  (2003). Located  near the  Southern Galactic Pole,  the observations
  for  the ESS  were  performed  as an  ESO  key-programme, thanks  to
  guaranteed time on the ESO NTT and 3.6m telescope.  Deep CCD Johnson
  $B$,  $V$ and  Cousins  $R_\mathrm{c}$ magnitudes  for nearly  13000
  galaxies to  $V \simeq24$  were obtained over  a continuous  area of
  $\sim0.37 \deg^2$ (Arnouts  et al.  1997). Multi-object spectroscopy
  has  also provided  redshifts  and flux  calibrated  spectra over  a
  sub-area of  $\sim0.25 \deg^2$  for 617 galaxies  with $R_\mathrm{c}
  \leq20.5$  ($92\%$  complete) and  870  galaxies with  $R_\mathrm{c}
  \leq21.5$ ($52\%$ complete).  The optical star/galaxy separation was
  performed using  the ``stellarity'' index from  the {\sc SExtractor}
  software  (Bertin  \&  Arnouts   1996).  The  optical  spectra  were
  classified  using  a  Principal  Component  Analysis  (Galaz  \&  de
  Lapparent 1998). The optical luminosity functions were then measured
  per galaxy spectral type (de  Lapparent et al. 2003), and lead to the
  detection of a  marked evolution in the spiral galaxy populations,
  characterised by an excess of  late  spiral  and
  irregular galaxies at $z\simeq 0.5-1.0$ (de Lapparent et al. 2004).

  %==========================================================
  \subsection{The \ISO survey on the ESS field (ISO-ESS)} 

  The \ISO observations of the ESS field have been performed with the 
  raster mode CAM01 of the \ISOCAM camera (Cesarsky et
  al.  1996) on board of  the {\it Infrared Space Observatory} \ISO (Kessler
  et al. 2003).  The broad-band LW10 filter, with reference wavelength
  $\lambda_{\text{ref}}=12\,\mu$m   and   covering   the   8.5--15.5$\,\mu$m
  interval, has been built for  a direct comparison with the 12$\,\mu$m
  $IRAS$\ filter  (Moneti et al. 1997).   The pixel field  of view (PFOV)
  with \ISOCAM is  a 6 arcsec square and  the adopted integration time
  was 5.04 sec. per exposure.  Ten  raster maps were  built within the
  total   on-target    time   of   14h.    All    the   rasters   have
  $M \times N=8 \times 8$\ pointings each offset by $dM=dN=60$\ arcsec along
  the axis  of the detector. The  number of exposures  per pointing is
  $N_{\text{exp}}=13$\   and  the   number  of   stabilization   exposures  is
  $N_{\text{stab}}=10$.  The total area of the survey is $\sim680$ arcmin$^2$
  with a  maximum exposure  time of 1200  sec. and an  average exposure
  time of $\sim660$ sec. per sky pointing.

  The  \ISO  target field  has  been selected  in  the  region of  the
  ESO-Sculptor  survey where  the  cirrus emission  is minimal  (see
  companion article).  The overlap  between  the 2  surveys  is  large, as  it
  represents  90\%  of  the  \ISOCAM  survey area,  and  75\%  of  the
  ESO-Sculptor spectroscopic area.

  %==========================================================
  \subsection{Summary of the data processing}

  This section  is a summary of  the data analysis  presented in the 
companion article
(Seymour et al. 2007).   The various  steps  of data  processing  
  are: (1)  deglitching
  (i.e. cosmic ray subtraction),  (2) correction of the long transient
  behavior etc.  performed by the PRETI software of Stark et al. (1999),
  and  (3) source  detection  above  a noise  map,  performed using  a
  wavelet  analysis technique.  The detailed  adaptation of  the PRETI
  method  to the  \ISOCAM data  analysis  was published  by Aussel  et
  al. (1999). We fine-tuned the  astrometry of the 12$\,\mu$m sources by
  comparison with  matched objects in the  2MASS catalogue, which
  led  to a  maximum  change of  0.3  arcsec in  the  position of  the
  12$\,\mu$m  sources.  The  subsequent  r.m.s.   offset  between  the
  \ISOCAM 12$\,\mu$m and the  ESS coordinates is 2 arcsec,  with no systematic
  offset.   The star/galaxy  separation  was performed  by a  detailed
  analysis  of  the  near-infrared  (NIR)  and  optical  colour-colour
  diagrams.  We then  derived  an empirical  flux  calibration from  a
  subset of stars in our sample. This calibration was based on fitting
  the optical-NIR counterparts of the stellar data with stellar spectrum
  templates  compiled  in the  P\'EGASE  library  and  then using  the
  optical/infrared  relations  from  IRAS  data to  predict  the  true
  infrared fluxes of stars in our field. This procedure took advantage
  of the deep photometric ESS survey in the optical.
% This negated the need for extensive simulations. 
  Our  flux calibration  is robust  as it  was carried  out  using two
  different  colour-colour relations  and empirical  calibration using
  well-known stellar templates.  The final catalogue of 142 sources 
  with \ISO fluxes is listed  in the companion article , and contains 
  22  stars and  120  galaxies. As expected, only a small fraction of 
the MIR sources are stars.

  %==========================================================
  \subsection{Large-scale distribution of the \ISOc-ESS sources}
 %______________________________________________ 
\begin{figure*}
\includegraphics[width=12cm]{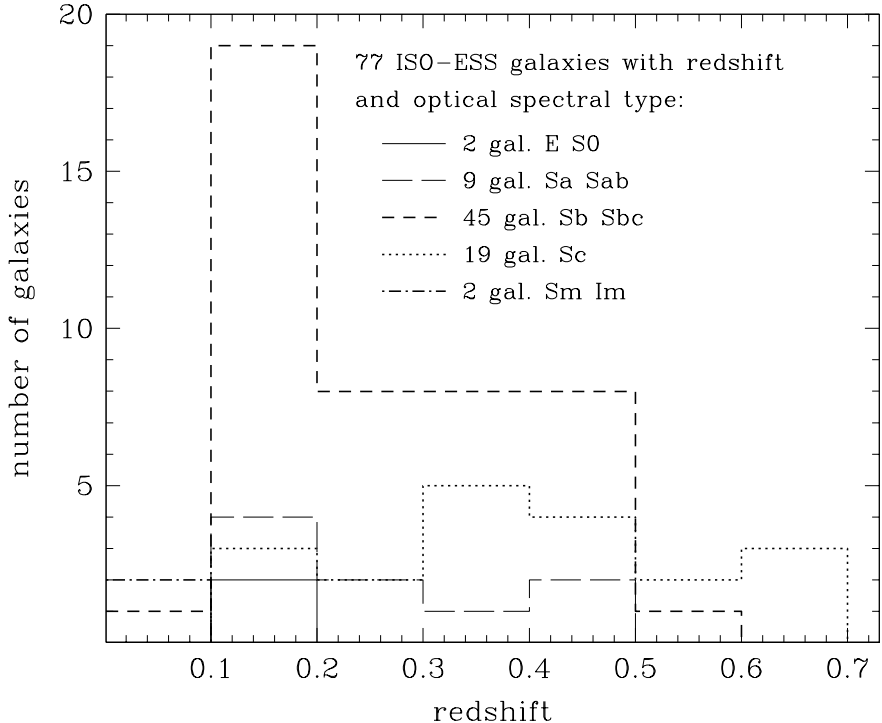}
        \caption{The  distribution by  types  of the  (R$<$20.5)
\ISOc-ESS galaxies,  derived from the  optical spectral classification
(de Lapparent et al. 2003). The dominant population is Sb-Sbc spirals,
confirming that  ellipticals are  rarely detected at  12$\,\mu$m. Note
however that among the 104 galaxies detected at $12\,\mu$m in the area
common  to both surveys,  only 77  have ESS  spectroscopic information
(hence redshift and spectral type), leaving place for the discovery of
new populations and signatures of evolution.}
      \label{figure:typezdist}
    \end{figure*}
% ______________________________________________ 
  With  an  average exposure  time  of  11  minutes by  pointing,  our
  \ISOc-ESS survey is as deep as the Lockman Hole Deep survey from the
  \ISOCAM Guaranteed  Time Extragalactic Survey (Elbaz  et al.  1999),
  but over a 33\% larger area of $\simeq$ 680 arcmin$^2$.  Despite its
  significant area, the \ISOc-ESS map reveals the inhomogeneity of the
  projected distribution of MIR sources.  The large-scale distribution
  of  galaxies in the  ESS shows  the alternation  of sharp  walls and
  voids  (Bellanger   \&  de   Lapparent  1995b).   For   $H_0$=65  km
  s$^{-1}$~Mpc$^{-1}$,  de  Lapparent   \&  Slezak  (2007)  measure  a
  comoving correlation length of  $\sim5.5$ Mpc at the median redshift
  $z\sim0.3$ of the \ISOc-ESS galaxies; the right ascension transverse
  extent   of  the   survey   ($\sim11$  Mpc)   thus  corresponds   to
  approximately twice the galaxy  correlation length.  Because the MIR
  sources  are likely to  follow the  clustering of  optical galaxies,
  large-scale fluctuations  are naturally expected.  These are however
  smoothed  out when  calculating the  \ISOc-ESS galaxy  counts summed
  over the whole redshift range of the survey.

  %%%%%%%%%%%%%%%%%%%%%%%%%%%%%%%%%%%%%%%%%%%%%%%%%%%%%%%%%%%

  \section{Faint galaxy counts at 12$\,\mu$m}
  
  The  \ISOc-ESS sample  consists of  sources with  a  detection level
  equivalent to 5$\sigma$  and is complete to 1  mJy.  At fainter flux
  densities,  the  correction   for  incompleteness  in  the  interval
  0.24--1.0 mJy  is computed  by two independent  methods respectively
  based on  stars and  galaxies in the  optical. Both  approaches take
  advantage of the deep ESS optical survey which contains counterparts
  to all sources  detected by \ISOCAM down to 0.24  mJy (in the common
  area to  both surveys),  and yield incompleteness  corrections which
  are in  good agreement  (see Fig. 7  in companion paper).   Here, we
  correct the galaxy number counts using the incompleteness correction
  derived from the  stars, as it is least  affected by the large-scale
  clustering    in    the     ESS    sample    (see    sect.     2.4).
  Fig.~\ref{figure:typezdist} shows  the distribution by  types of the
  77 \ISOc-ESS galaxies which have measured redshifts from the optical
  survey (de Lapparent et al. 2003).  It shows that spirals Sb and Sbc
  are the most numerous galaxies detected in the MIR. It also confirms
  that gas-poor normal ellipticals are essentially undetected. The ESS
  spectral   classification  used   the  code   P\'EGASE.2   (Fioc  \&
  Rocca-Volmerange     1997).       However,     the     sample     on
  Fig.~\ref{figure:typezdist}  is incomplete,  with  only 77  galaxies
  having a measured redshift.  Among  the 104 ESS galaxies detected at
  12$\,\mu$m, the 27 objects with  no redshift measurement could be at
  higher  redshift or belong  to a  new type.  We therefore  adopt the
  ``optical'' density fractions derived  from the faint count analysis
  in the UV-optical-nearIR ($\simeq$  27\%, 30\%, and 43\% for early-,
  intermediate-, and  late-type galaxies respectively) by  Fioc et al.
  (1999a), confirmed  by de  Lapparent et al.  (2003) in  the Sculptor
  field.

  %==========================================================
  \subsection{Cumulative galaxy number counts N($\geq S_{\nu}$) at 12$\,\mu$m} 

  %______________________________________________ 
  \begin{figure*}
\includegraphics[width=15cm]{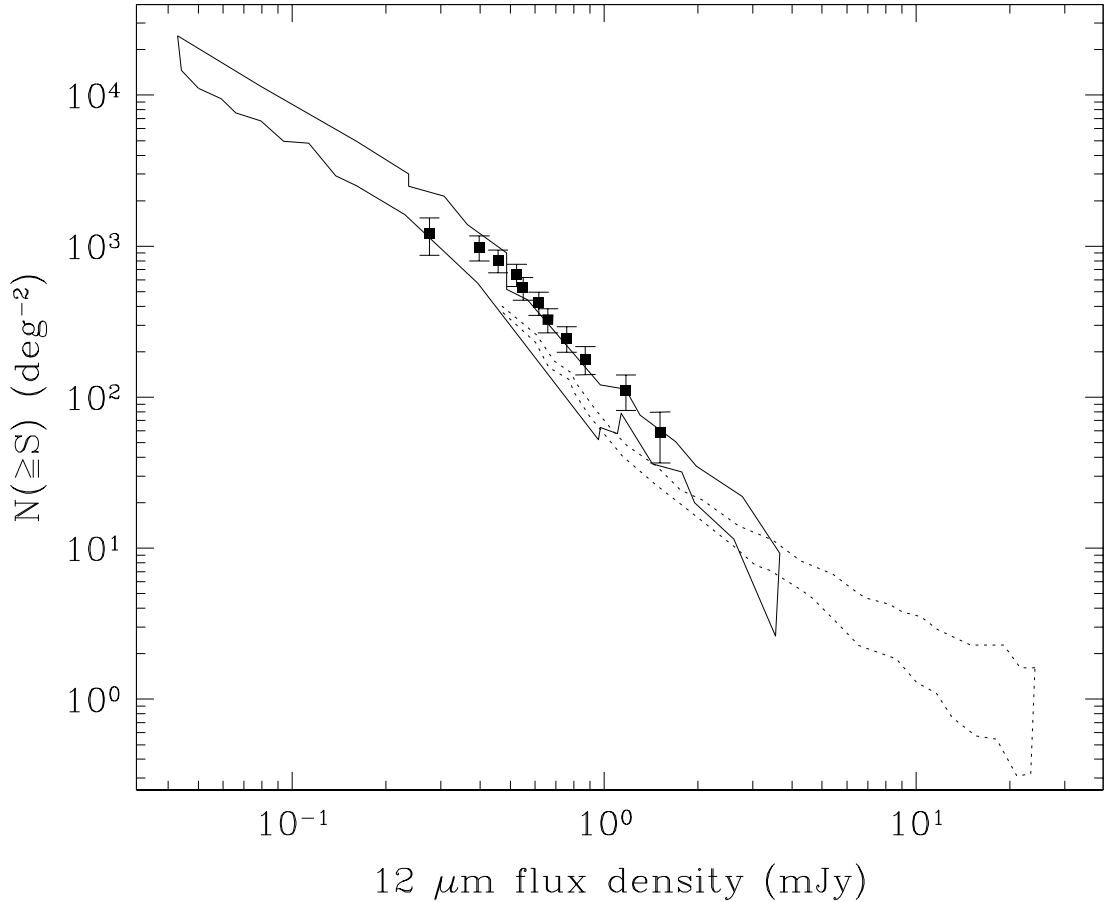}
    \caption{Cumulative  12$\,\mu$m galaxy  counts observed 
with the LW10 filter from  the \ISOc-ESS
survey,  plotted as  full  squares  with error  bars. The  other
\ISOCAM 15$\,\mu$m surveys are  also plotted: the compilation by Elbaz
et al. (1999)  as a solid line contour, which  covers the Lockman Hole
(Rodighiero et al. 2004), the  Marano field (Aussel et al.  1999), and
the  HDF North  and  South fields  (Altieri  et al.  1999) to  various
depths;  the  number-counts  from  the  ELAIS  survey  (La  Franca  et
al. 2004)  are shown as  a dotted line.  The 15$\,\mu$m counts are
corrected to 12$\,\mu$m, see text for details.}
      \label{figure:Int_nomodel}
  \end{figure*}
  %______________________________________________ 

  We then derive the number of detected \ISOc-ESS galaxies as a function
  of 12$\,\mu$m  flux density.  The detections are  binned in  11 flux
  density intervals, chosen  so that each bin contains 
  $10-12$ more sources than the previous one.  The resulting cumulative source counts $N(\geq S_{\nu})$
  (commonly referred  to ``integrated counts'' in  the literature), where
  $S_\nu$ is the 12$\,\mu$m flux  in mJy, derived after correction for
  incompleteness, are  shown in Fig.~\ref{figure:Int_nomodel}  as full
  squares.  The vertical error bars in the corrected cumulative counts
  are estimated  by assuming Poisson fluctuations in the number  of detected
  galaxies,  and  by  taking  into  account  the  uncertainty  in  the
  incompleteness correction; the horizontal error bars are not plotted
  as they are  smaller than the size of the  symbols.  The large error
  bars in the high flux  point at $S_\nu\simeq1.4$ mJy result from the
  small number of galaxies (10) detected at these high flux densities:
  this is evidently due to the limited area of the \ISOc-ESS field.

  We  overlay in  Fig.~\ref{figure:Int_nomodel} the  results  from the
  compilation of \ISOCAM 15$\,\mu$m surveys obtained in the LW3 filter
  as  published by Elbaz  et al.  (1999; solid  line): they  cover the
  Lockman Hole, Marano  and HDF North and South  fields (Rodighiero et
  al.  2004; Aussel  et al.  1999; Altieri  et al.   1999)  at various
  depths.   We also  plot for  comparison the  results from  the ELAIS
  survey, also observed at 15$\,\mu$m  (La Franca et al.  2004, dotted
  line). When plotting the  15$\,\mu$m number counts, the flux density
  is  converted into  the LW10  12$\,\mu$m band  using  the respective
  central wavelengths of the two  filters and assuming a flat spectrum
  in flux  (that is, a  constant product of  the frequency by  the flux
  density); the correction  is small, a factor $12/15=0.8$.  We do not
  correct the counts from 15$\,\mu$m to 12$\,\mu$m because the filters
  are wide and  very close to each other  in wavelength (they actually
  slightly overlap).

  Figure~\ref{figure:Int_nomodel} shows that our data are in good  
  agreement  with  the  surveys  compiled  by  Elbaz  et  al.  (1999),
  including the  deepest samples.  In particular,  the faint-end slope
  of the  \ISOc-ESS cumulative  counts is similar  to that of  the ultra
  deep  survey  obtained  on  the  cluster-lens A2390  by  Altieri  et
  al. (1999) in their common flux density interval, despite the narrow
  pencil-beam geometry  of the latter survey.  The  \ISOc-ESS counts are
  also consistent  with the brighter  galaxy survey from La  Franca et
  al. (2004).

  %==========================================================
  \subsection{The Euclidean-normalized differential counts}

  %______________________________________________ 
    \begin{figure*}
\includegraphics[width=15cm]{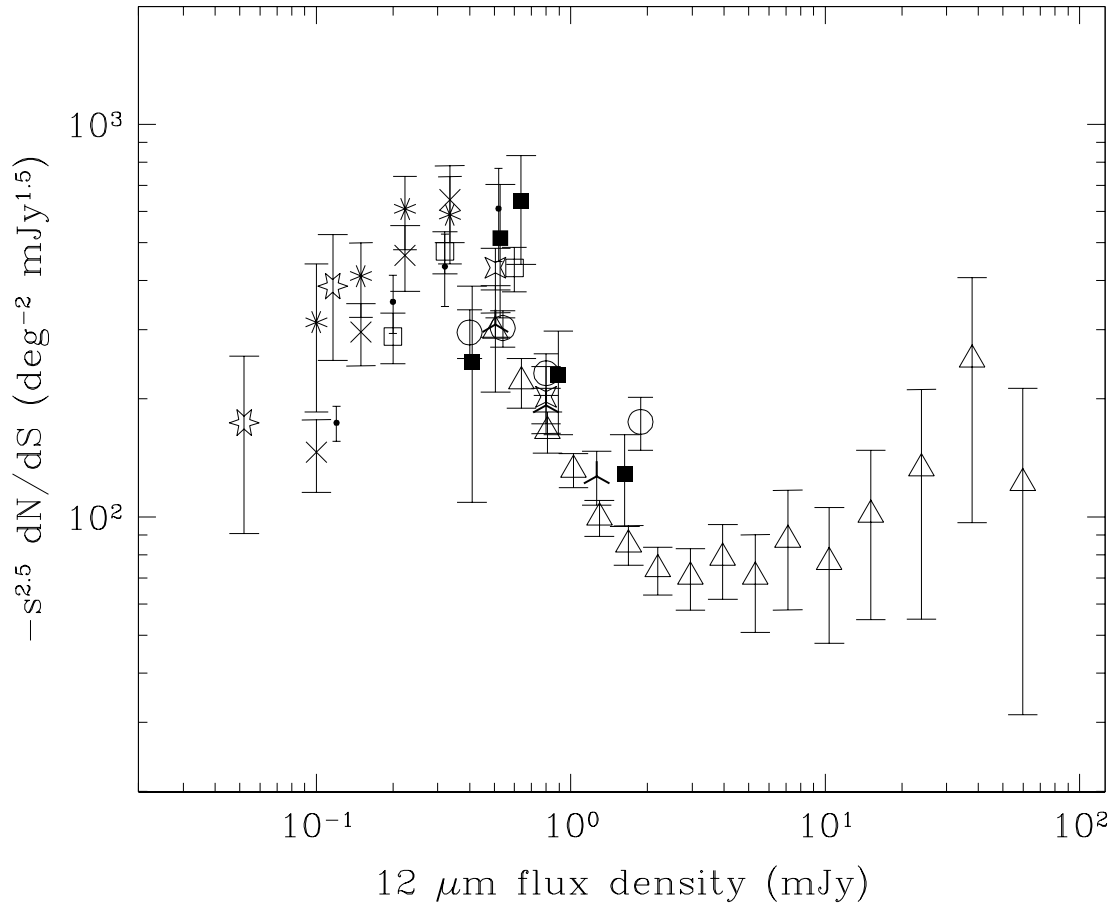}
      \caption{ The \ISOc-ESS  12$\,\mu$m differential counts normalized
to the Euclidean case, plotted as filled squares  (note that -$S^{2.5}
dN/dS$\ is positive).  The  other following 15$\,\mu$m surveys are also
shown:  ELAIS-S1  (Pozzi  et  al.   2003,  triangles);  through  the
cluster-lens  A2390 (Altieri et  al. 1999,  6-pointed open  star), HDF
North (Aussel  et al.  1999,  crosses); HDF South  (asterisks), Marano
FIRBACK Ultra Deep field (open squares), Marano Ultra Deep ROSAT field
(points), Marano Deep field (open  circles), all published in Elbaz et
al.  (1999); Lockman Deep field (4-pointed open star), Lockman Shallow
field (3-pointed heavy cross), both from Rodighiero et al. (2004). See
text for the details of the correction from 15$\,\mu$m to 12$\,\mu$m.}
      \label{figure:Diff_nomodel}
    \end{figure*}
  %______________________________________________ 

The differential galaxy counts  -$S^{2.5} dN/dS$($>$0 ) of the \ISOc-ESS
survey  at  12$\,\mu$m,  normalized   to those for an  Euclidean  Universe  are
presented as filled squares in Fig.~\ref{figure:Diff_nomodel}. A wider
binning than in Fig.~\ref{figure:Int_nomodel} is adopted, with $23-25$
galaxies  per bin.  The  vertical  error bars  take  into account  the
Poisson  error in the  number of  actually detected  sources, combined
with  the uncertainties in  the incompleteness  correction and  in the
flux density of each detected source; as for the cumulative counts the
horizontal error  bars are  not plotted as  they are smaller  than the
size of the  symbols.  The Euclidean normalization is  adopted so that
for a static universe with  a non-evolving population of objects and a
constant luminosity  function, the Euclidean  normalized galaxy counts
would follow a horizontal line.

%  are chosen to be of equal size in log flux density, but the
%  brightest 3 bins are combined to improve the statistics.  The number
%  of sources in each bin is indicated by the number below it on the
%  diagram.

  The differential number counts  from the surveys at 15$\,\mu$m (with
flux density  converted into the 12$\,\mu$m band,  see previous Sect.)
are also plotted in Fig.~\ref{figure:Diff_nomodel} (ELAIS-S1, Pozzi et
al. 2003; A2390, Altieri et al.  1999; HDF North, Aussel et al.  1999;
HDF South,  Marano FIRBACK Ultra  Deep field, Marano Ultra  Deep ROSAT
field, Marano Deep field, Elbaz et al.  1999; Lockman Deep and Lockman
Shallow field, Rodighiero et al.  2004). As for cumulative counts, the
15$\,\mu$m  counts are  corrected  to 12$\,\mu$m  (see Sect{.}   3.1).
Note that our faintest bin (at $\sim0.3$ mJy) is very uncertain due to
incompleteness,  which may explain  the downward  shift for  this last
point.  For  clarity we  do not include  the 12$\,\mu$m  galaxy counts
from Clements et  al. (1999) which show a  large scatter, probably due
to poor statistics  (3--5 objects per bin). They  may also suffer from
contamination by a few stars at the bright end: the authors admit that
there  is  at least  one  star in  their  galaxy  counts (Clements  et
al. 2001).
  
  In  Fig.~\ref{figure:Diff_nomodel},  the  \ISOc-ESS  counts  show  a
  strong departure from Euclidean  no-evolution models at faint fluxes
  ($<1$  mJy), with  a  very steep  super-Euclidean  slope.  The  same
  behavior is observed in all the other surveys plotted in the figure,
  showing  agreement  among  the   15$\,\mu$m  surveys  and  with  our
  12$\,\mu$m survey.  The low value  of our last point at faint fluxes
  ($\sim0.3$   mJy)  is   due  to   the  large   uncertainty   in  the
  incompleteness correction.   In the following, we  address the issue
  of  the origin of  the excess  counts by  modeling the  faint galaxy
  counts with the evolutionary code P\'EGASE.

  %%%%%%%%%%%%%%%%%%%%%%%%%%%%%%%%%%%%%%%%%%%%%%%%%%%%%%%%%%%

  \section{Modeling MIR galaxy counts per type with the code P\'EGASE.3}
 In the following we  adopt a flat
Universe  with  the  standard  cosmological  parameters:  $H_0=72$  km
s$^{-1}$   Mpc$   ^{-1}$,   $\Omega_M=0.3$,   $\Omega_{\Lambda}=0.7$
(Spergel et al. 2003).

   %===========================================         
       
   \subsection {The code P\'EGASE.3}

   %______________________________________________ 
    \begin{figure*}
\includegraphics[width=10cm,angle=-90]{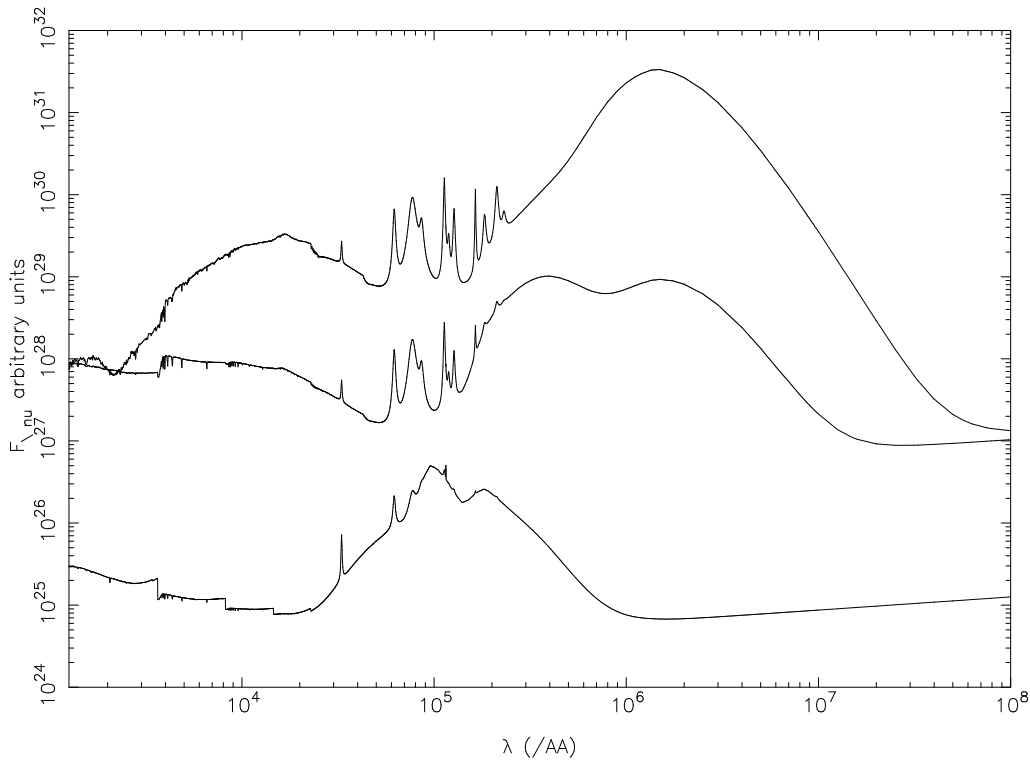}
\caption{Example of the strong evolution in the SEDs predicted 
by P\'EGASE.3  
for a star forming galaxy  spiral Sc at various ages 
(increasing upwards), in the wavelength interval 1000 \AA\ to 1 cm.
(see more details in Fioc et al. 2007)}
     \label{figure:spirals} 
    \end{figure*} 
   %______________________________________________ 

The  new  evolutionary code  P\'EGASE.3  (Fioc  et  al.  2007)  is  an
extension of the code P\'EGASE.2 to the dust emission wavelength range
(Fioc     \&     Rocca-Volmerange     1997,    1999b,     see     also
http://www2.iap.fr/pegase).   The  SEDs   of  reddened   galaxies  are
consistently   computed  from  the   far-UV-optical-NIR  to   the  FIR
(far-infrared) domain.   P\'EGASE.3 calculates, in  a coherent manner,
the stellar emission, extinction,  metal-enrichment, dust mass and the
emission of  grains statistically heated by the  radiation field.  Two
distinct  dust  media  (interstellar   medium  and  HII  regions)  are
considered.  As  in P\'EGASE.2, radiation transfer is  computed in two
geometries  (slab  and spheroid)  appropriate  for  disk galaxies  and
ellipticals.  Temperature  fluctuations  of  the  Polycyclic  Aromatic
Hydrocarbons (PAH)  as well as graphite and  silicate grain properties
are derived  with the  method of Guhathakurta  \& Draine,  (1989). For
illustration, Fig.~\ref{figure:spirals} shows  the strong evolution of
the spectral  energy distribution  (SED) of the  spiral Sc  template
(specially  in the MIR  between 10$\,\mu$m  and 20$\,\mu$m)  at various
ages.

   %===========================================         
 
  \subsection {Evolutionary scenarios of galaxies and relative 
number density fractions}
\subsubsection{Model 1: "normal" evolved types}
In a  first step, we adopt  the same set of  evolutionary scenarios of
``normal''  galaxies previously  determined with  the  code P\'EGASE.2
(Fioc \& Rocca-Volmerange 1997) which fit the colors of nearby
galaxies by types and the deepest multi-wavelength ($B_J$, $U$ (and F300W), $I$
and $K$)  faint galaxy  counts in the  UV-optical-NIR ranges  (Fioc \&
Rocca-Volmerange  1999a).  This  set  corresponds to  the 8  following
types: irregular magellanic Im; spirals  Sd, Sc, Sbc, Sb, Sa; and
ellipticals E2  and E). We  use the same  parameter set with  the new
version   P\'EGASE.3  (Fig.~\ref{figure:spirals})   and   compute  the
continous SEDs of  each type, extended to the  mid- and far-IR, taking
into account  stellar and dust  emissions as well as coherent absorption. 
The initial mass function is from  Rana \& Basu (1992) for each type.
These templates are  also able to predict photometric  redshifts up to
$z$=4  with   an  accuracy  $\sigma_z  \simeq  0.1$\   (Le  Borgne  \&
Rocca-Volmerange  2002).    Therefore  the  evolution   scenarios  are
considered  as  robust.   The  main parameters  (star  formation  law,
initial mass  function, galactic winds and astration  rate) are listed
for each  type in  Fioc \& Rocca-Volmerange  (1997) and Le  Borgne \&
Rocca-Volmerange (2002).  Star formation rates are  proportional to the
current  gas  mass  density,  a  highly  conservative assumption. The
astration parameter  $p_2^{-1}$ varies  with galaxy type.  The current
gas content  $M_{\text{gas}}(t)$ is ruled by star  formation, stellar ejecta,
galactic   winds  and   infall   rates  as   described   in  Fioc   \&
Rocca-Volmerange (1997); the adopted values by types are recalled in Table 1.

In   Table    1,   we   present    the   characteristic   luminosities
 $\log(L_*(12\,\mu\mathrm{m})/L_\odot)$\                            and
 $\log(L_*(24\,\mu\mathrm{m})/L_\odot)$\   adopted  for   the  various
 galaxy    types.     We    compute    $L_*(12\,\mu\mathrm{m}$)    and
 $L_*(24\,\mu\mathrm{m}$)  from the $L_*(B_J)$  values of  the optical
 luminosity functions  by types (Heyl et  al.  1997), used  to fit the
 faint optical counts (see Table 1 of Fioc \& Rocca-Volmerange 1999a),
 and the colours $B_J-12\,\mu$m  and $B_J-24\,\mu$m from P\'EGASE.3 at
 $z=0$ for each galaxy type (Fioc et al. 2007).  The filter 12$\,\mu$m
 means \ISOc/LW10 and  24$\,\mu$m means \Spitzerc/MIPS 24$\,\mu$m; $B_J$ is
 the blue Kodack  IIIa-J plus GG395 corrected filter  (Couch \& Newell
 1980).  The filter corrections from $IRAS/12\,\mu$m to \ISOc/LW10 and
 from $IRAS/25\,\mu$m  to \Spitzerc/24$\,\mu$m are  taken into account
 by the  code.  Among normal  galaxies at $z\simeq0$, spirals  Sbc are
 the  brightest emitters at  12$\,\mu$m and  24$\,\mu$m, and  also the
 most numerous (see Fig.~\ref{figure:typezdist}). 
 We therefore assign
 to      type       Sbc      the      characteristic      luminosities
 $\log(L_*(12\,\mu\mathrm{m})/L_\odot)=9.8$\                         and
 $\log(L_*(24\,\mu\mathrm{m})/L_\odot)=9.6$\ derived from the observed
 $z\simeq0$ luminosity  functions measured by Rush et  al.  (1993) and
 by  Shupe  et  al.   (1998)  respectively.  We  then  scale  the  MIR
 luminosities of all the other types accordingly.

Because  the evolutionary  scenario of  ellipticals (see  Fig.   3a of
Rocca-Volmerange  et   al.   2004)  may play  a  specific   role  in  the
interpretation of observations of the ultra luminous infrared galaxies
(ULIRGs), defined as galaxies with infra red luminosities 
$\geq$10$^{12}$L$_{\odot}$, it deserves more  attention. The intense star formation rate
(low $p_2$\ value) in  the first Gyrs  is fueled  by a  high  
infall rate  from the  gas
reservoir.  The activity is so intense that a huge dust mass is formed
at  early  epochs  from  stellar  ejecta,  specifically  from  massive
supernovae  SNII. In  normal elliptical  galaxies, the  star formation
activity is supposed to be  halted when strong galactic winds produced
by the  bulk of  supernovae expel  all the gas  and dust  content from
galaxy.  Most  of stars and dust  are already formed  when the galaxy
age is  of about 1 Gyr, corresponding to  $z\simeq 4$ in  the adopted
cosmology. ``Normal''  ellipticals contribute very  little thereafter to
the infrared emission  as they are largely dust-free  from this age to
the  present time  (0 $\leq  z \leq  $4).  This  scenario at $z$=0 matches the
observation that the cold  grain component ($\sim$\ 50K) in elliptical
galaxies has almost  no contribution to the MIR  flux (Xilouris et al.
2004).

Column  7 of Table  1 gives  the relative  number fractions  of galaxy
types for model 1, as  derived from the UV-optical-nearIR.  This model
has no  ``ULIRG''component and is  thus composed only of  normal types
(26.5\% ellipticals (E + E2), 23.7\%  Sa to Sbc spirals, 33.2\% Sc, Sd
spirals and  16.6\% irregulars)  which were found  to fit  the deepest
UV-optical-nearIR  faint  galaxy  counts  (see  Fig.~5  from  Fioc  \&
Rocca-Volmerange 1999a).

\subsubsection{Model 2: dusty ellipticals as ULIRGs}
  Column 8  of Table 1 describes  our model 2 with  ``ULIRG'' which we
use  to adjust  the MIR  galaxy  counts.  In  this model,  1/3 of  the
ellipticals   have    over-luminous   MIR   luminosities    given   by
$\log(L_*(12\,\mu\mathrm{m})/L_\odot)=9.9$                          and
$\log(L_*(24\,\mu\mathrm{m})/L_\odot)=9.7$,  which  correspond to  the
observed characteristic $L*$ of  the MIR luminosity functions at these
two wavelengths. At  $z=0$\ and in the MIR, they  are nearly as bright
as Sbc spirals, and respectively brighter by 2.5 mag at 12$\,\mu$m and
5 mag at 24$\,\mu$m than  normal ellipticals, whatever their type (E2,
E).  However, they better  follow the evolution scenario of elliptical
galaxies of  type E, with  the same astration rate  $p_2^{-1}$, infall
and galactic  winds at the same age. These overluminous ellipticals,
forming at  early epochs large masses  of dust and  stars, become much
brigther  at high  $z$ than  spirals.  This  model does  not  need any
additional starburst, as  that seen in the typical  case of M82 (Silva
et al., 1998). Our model remains compatible with occasional starbursts
of  short  time-duration  ($<  10^8$  yrs), concerning  a  small  mass
fraction relative to the massive underlying elliptical galaxy.

Column 8  of Table 1 lists  the number fractions for  model 2.  The
population of ULIRG/dusty massive ellipticals correspond to 9\% of the
total number of galaxies.  The number of normal
dust-poor ellipticals is then reduced  to only 17.5\% (only 2/3 of the
ellipticals observed in the  visible). 
The rest of galaxies are normal; model 2 
therefore respects the majority of fractions by type derived from the
UV-optical-nearIR galaxy counts.

\subsubsection {The $k(z)$ and $e(z)$ corrections per type}
To  calculate the apparent magnitudes at  high $z$, the evolutionary
$e(z)$  and cosmological  $ k(z)$  corrections are  computed  for each
type,  as in  Rocca-Volmerange \&  Guiderdoni (1988),  and are
applied to the $z=0$ SEDs:
\begin{equation} m(z,t_z) =  M(0,t_0) + k(z) + e(z) + 5\log_{10}[D_A (1+z)^2] +
25 \label{eq:magapp}
\end{equation}
where  $D_A$  is the  angular  diameter  distance in Mpc,  $D_A (1+z)^2$  the
luminosity  distance, $t_z$\  and $t_0$\  the  cosmic times  at $z$\  and
$z=0$; internal extinction is taken  into account in our scenarios; no
Galactic extinction term is added, because our deep survey is made
in a region of weak galactic absorption.

%%%%%%%%%%%%%%%%%%%%%%%%%%%%%%%%%%%%%%%%%%%%%%%%%%%%%%%%%%%
\begin{table}[!htbp]
\begin{center}
\begin{tabular}{|c|c|c|c|c|c|c|c}
\hline
Type &$p_2$&$log_{10}$&$log_{10}$ & Colour& Colour&Model 1 &Model 2  \\ 
 &(Myr)&$[L_{*}(12\,\mu\mathrm{m})$& [$L_{*}(24\,\mu\mathrm{m})$&$B_J-12\,\mu$m&$B_J-24\,\mu$m&normal& normal\\
  & &/$L_{\odot}$] &/$L_{\odot}]$ & $_{AB}$& $_{AB}$& only & + ULIRG \\
\hline \hline
ULIRG&100&9.9&9.7&$-$&$-$&0\%& 9.0\%\\ \hline
E&100&8.9&7.7& -1.00&-2.60&9\%& 0\%\\ \hline
E2 &300&9.0&7.8&-1.20&-2.82&17.5\%&17.5\%\\ \hline
Sa &1400&9.0&8.8& -0.45&+0.21& 7.9\%& 7.9\%\\ \hline
Sb &2500&9.4&9.2&+0.37&+1.28& 7.9\%& 7.9\% \\ \hline
Sbc &5714&9.8&9.6&+1.38&+2.30&7.9\%& 7.9\%\\ \hline
Sc &10000&9.6&9.4&+1.57&+2.46&16.6\% & 16.6\%\\ \hline
Sd &14286&9.6&9.4&+1.61&+2.48&16.6\%& 16.6\%\\ \hline
Im &16000&8.8&8.5&+1.57&+2.42&16.6\%& 16.6\%\\ \hline
\end{tabular}
\caption{Characteristic luminosity  $L_{*}$ in units  of $L_\odot$\ at
12$\,\mu$m  (Column 3)  and 24$\,\mu$m  (Column 4)  of  the luminosity
functions used here, as a  function of galaxy type (Column 1).  Column
2  gives  the  value of  the  parameter  $p_2$,  the rate  defined  by
$SFR_i(t)=M_{\text{gas}}(t)/{p_2}$.  Galactic winds occur at an age of
3 Gyr in ellipticals of type E  and ULIRG, and at 1 Gyr in ellipticals
of type E2; there are no galactic winds in spirals.  Infall time-scale
of ULIRGs is 100 Myr as for  ellipticals of type E or E2; it regularly
increases for spirals from 2.8 Gyr  (Sa) to 8.0 Gyr (Im) (see text for
details and references).  The $L_{*}$(12$\,\mu$m) for the normal types
(other than ULIRG) are derived  from the $L_{*}(B_J)$\ of the observed
optical  luminosity  functions  from   Heyl  et  al.  (1997)  and  the
$B_J-12\,\mu$m  and  $B_J-24\,\mu$m  colours,  in the  $AB$  magnitude
system, are computed at $z\simeq0$  with P\'EGASE.3 (Columns 5 and 6).
The value of $L_{*}(12\,\mu$m) assigned to Sbc galaxies, the brightest
emitters in  this band,  is that of  the observed  luminosity function
measured with $IRAS$  (Rush et al. 1993; Shupe et  al. 1998); the same
offset as for Sbc is then  applied to all types.  The last two columns
show  the adopted  number density  fraction by  type: model  1 (normal
galaxies only, column  7) uses the type distribution  derived from the
UV-optical-nearIR  faint counts  (Fioc et  al.  1999a)  while  model 2
(normal  galaxies +  ULIRG, column  8) is  built by  replacing  1/3 of
normal dust-poor ellipticals (9\% of all galaxies) with ultra luminous
elliptical  galaxies (called  ULIRGs). These  last galaxies  evolve as
dusty ellipticals, they are $\simeq  2.5$\ to $\simeq 5$\ mag brighter
than normal  ellipticals (depending on  wavelength) , and are  thus as
luminous as spirals Sbc at  $z=0$.  The number densities for the other
galaxy types are identical in both models. }
\end{center}
\label{table:parameters} 
\end{table}
%%%%%%%%%%%%%%%%%%%%%%%%%%%%%%%%%%%%%%%%%%%%%%%%%%%%%%%%%%%%%%%%%%%%%%%%%
 The same  formation redshift  of $z_{\text{for}}=10$ is  arbitrarily adopted
for all galaxy  types, following the most distant
galaxies discovered at $z>6$ (Hu et al. 2002; Chary et al. 2005).
%%%%%%%%%%%%%%%%%%%%%%%%%%%%%%%%%%%%%%%%%%%%%%%%%%%%%%%%%%%%%%%%%%%%%%%%%%

  \subsection{The 12$\,\mu$m luminosity function by type at z=0}

We  take  advantage  of   the  quasi-similarity  of  the  two  filters
\emph{IRAS}/12$\,\mu$m  and  {\ISOCAMc/LW10/12$\,\mu$m}  (the  flux
differences are $<\ 10\%$,  see companion article) and use  the local 
12$\,\mu$m
luminosity  function measured  by Rush  et al.   (1993) from  the $IRAS$\
catalogue  (this  measurement  was  later confirmed  and  extended  to
fainter fluxes by Fang et al. 1998),
\begin{equation}
\phi_{0}(L)=C\left(\frac{L}{L_*}\right)^{1-\alpha}\left(1+\frac{L}{\beta
L_*}\right)^{-\beta}.
\label{eq:lf}
\end{equation} 
We adopt the values $\alpha=1.7$, $\beta= 3.6$\ measured by Rush et al.
(1993) for the non-Seyfert  galaxies, which include \emph{all} galaxy
types detected at $z=0$.
As shown in Table 1, the  $L_{*}(12\,\mu$m) adopted for each galaxy type
  are derived 
by combining  the optical $L_{*}(B_J)$  (Heyl  et  al. 1997) with the
$B_J$-12$\,\mu$m  and  $B_J$-24$\,\mu$m  colors predicted by 
the code P\'EGASE.3 (columns 5  and 6).  We assign to Sbc
galaxies, the brightest  emitters in these bands, the $IRAS$\ 
values of  $L_{*}$(12$\,\mu$m) from the
observed luminosity  functions (Rush  et al. 1993,
Fang et al, 1998). The same offset  as for Sbc is then applied to all
types.

  \section{Results of galaxy count modeling at 12$\,\mu$m} 
   %===========================================         
   \subsection {Cumulative and differential counts from P\'EGASE.3 at $12\mu$m}
%______________________________________________ 
     \begin{figure*}                
\includegraphics[width=7.5cm]{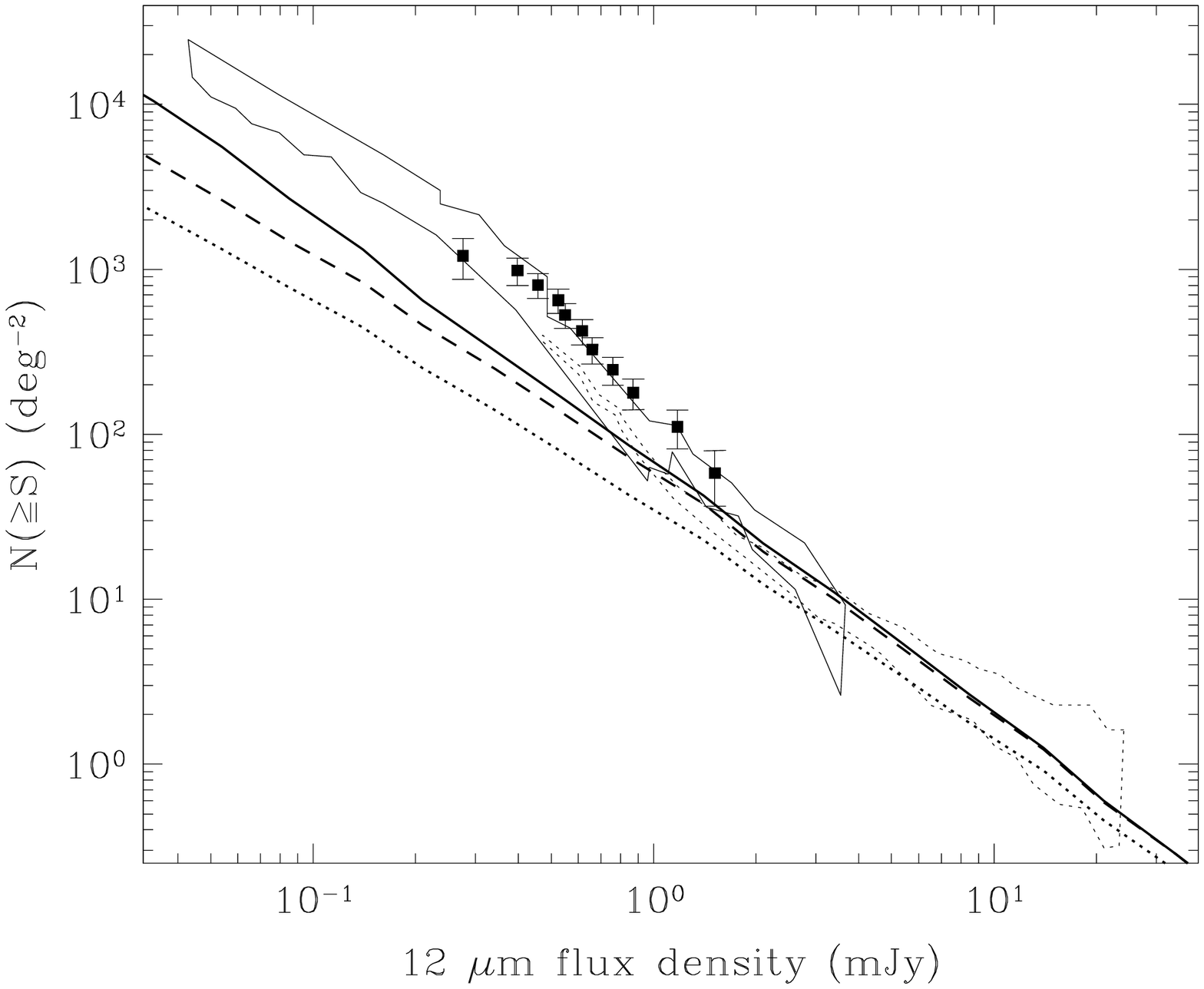}
\includegraphics[width=7.5cm]{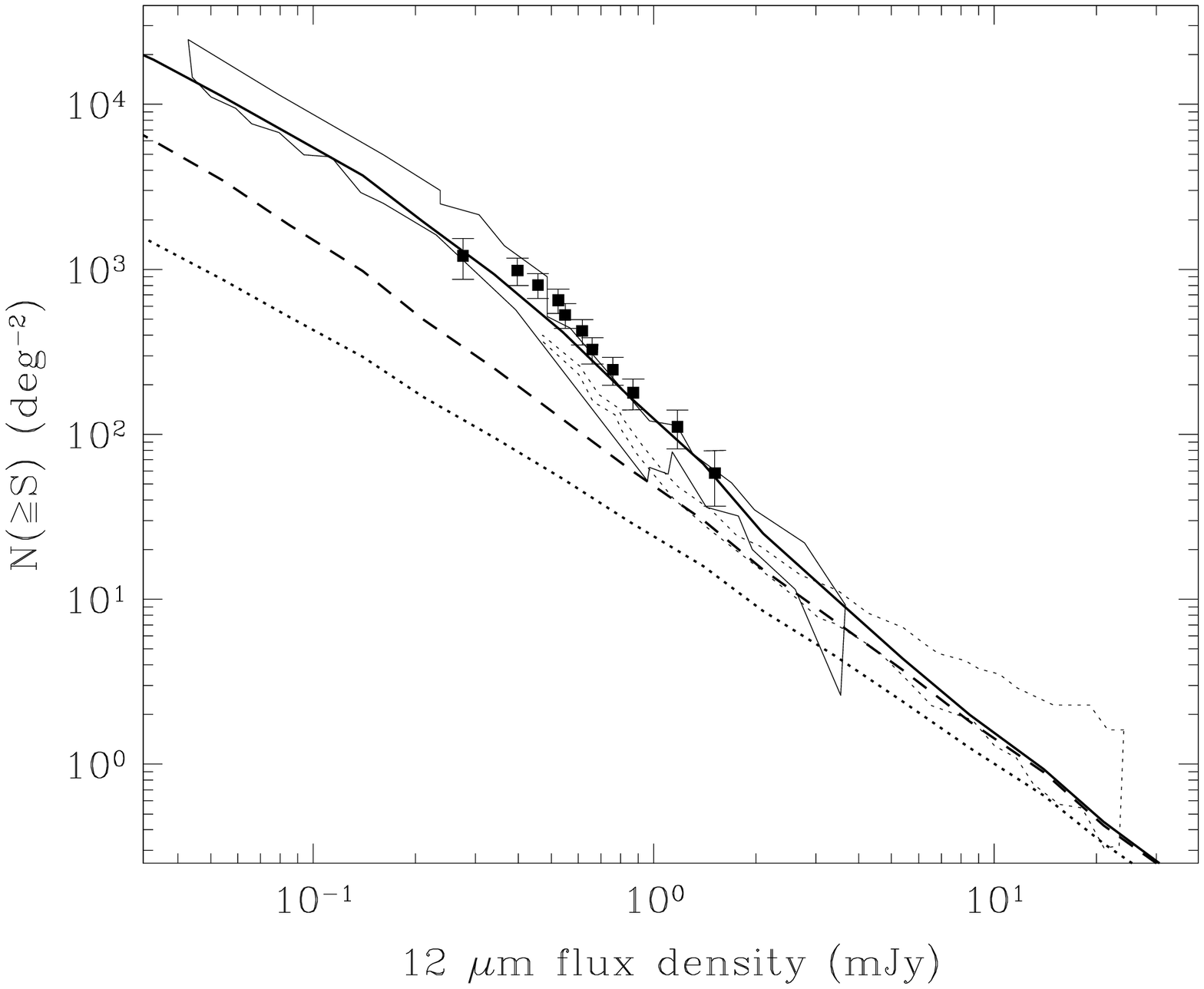}
\caption{The  predictions of 12$\,\mu$m cumulative
  galaxy  counts (heavy line) calculated from model 1 of 
normal galaxies (left) and model 2 with 9\% of ULIRGs (right):  
see Table 1, columns 7 and 8). They
       are compared  to the  observed 12$\,\mu$m cumulative counts from 
the  \ISOc-ESS survey (black squares).  
      The models with the cosmological
       $k$- correction only (no evolution correction: heavy dashed line) 
      and the models with only the elemental comoving volume effect
      (dotted  line) are  also plotted for comparison. 
	 The  faint counts from  the
       15 $\,\mu$m surveys are shown with the same line coding
       as  in Fig.~\ref{figure:Int_nomodel}.
}   
       \label{figure:Int}   
     \end{figure*}
   %______________________________________________
    \begin{figure*}  
\includegraphics[width=7.5cm]{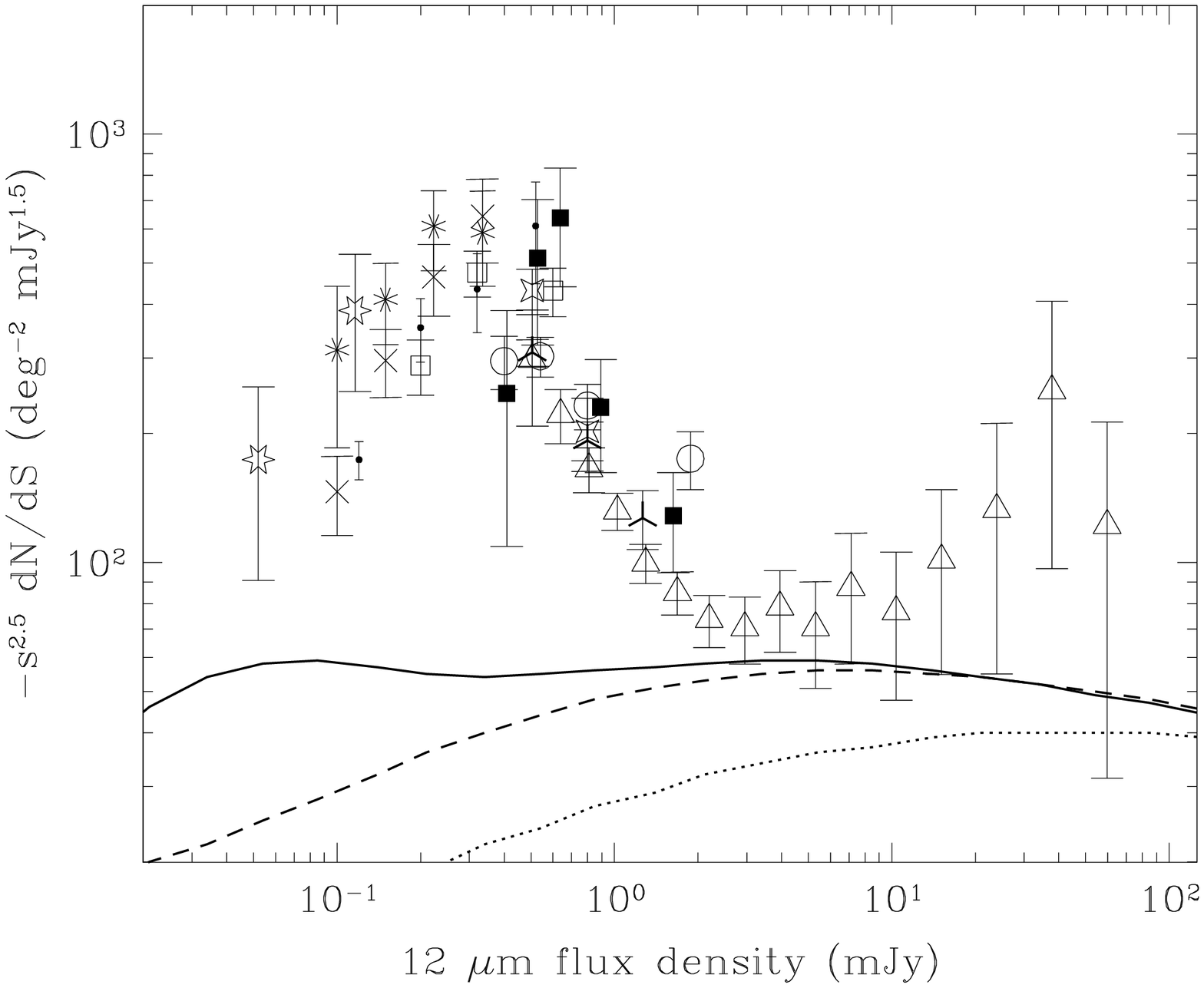}
\includegraphics[width=7.5cm]{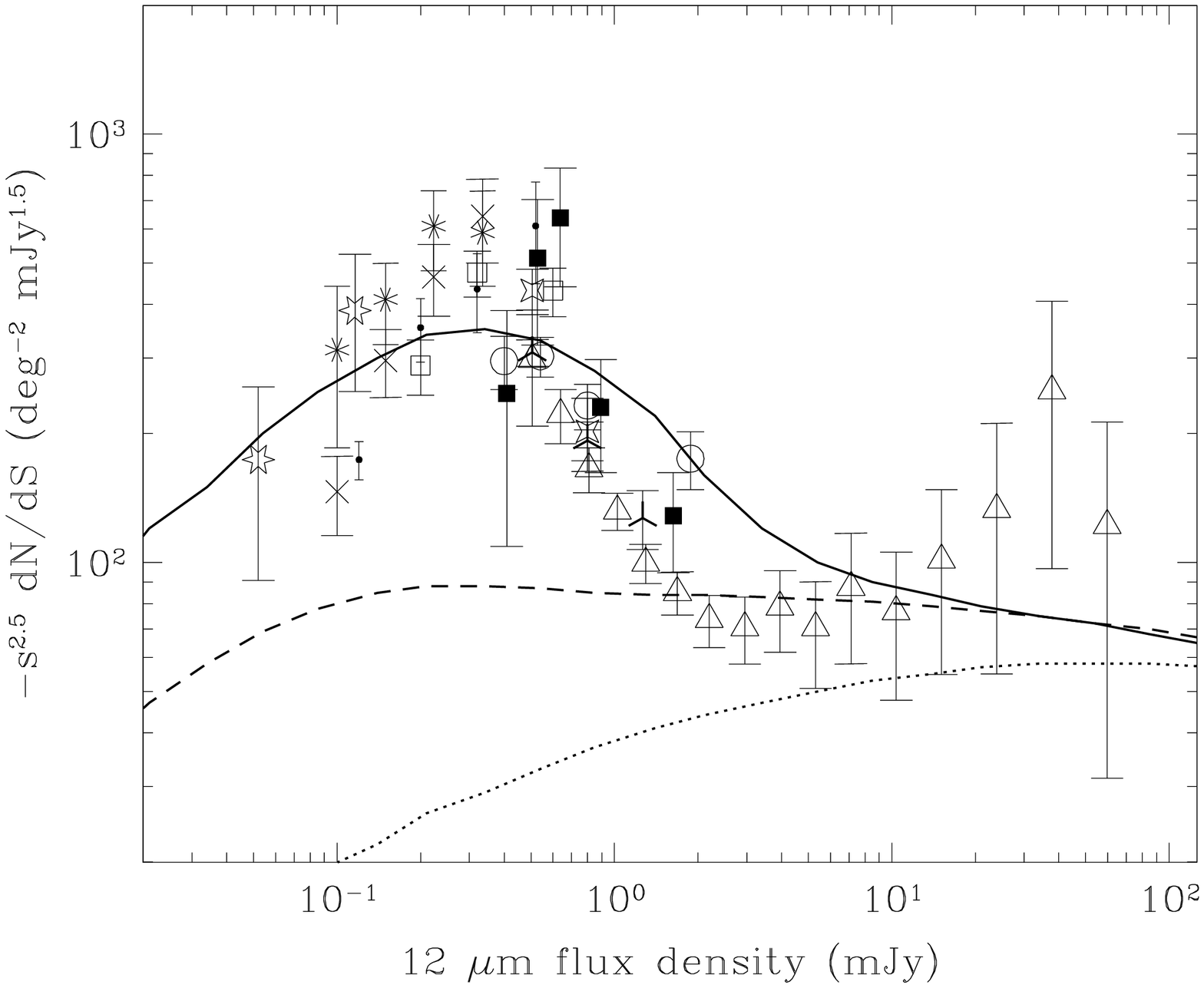}
    \caption{The  predictions   of  Euclidean-normalized  differential
         12$\,\mu$m  galaxy counts, calculated  with model  1 (normal,
         left)  and  model 2  (with  ULIRGs,  right) using  P\'EGASE.3
         (solid  line),  are  compared  to the  observed  differential
         counts  at  12$\,\mu$m   from  the  $ISO$-ESS  survey  (black
         squares). The cases without evolution (only cosmological $k$-
         correction) are  shown with a  dashed line, the  case without
         $k+e$ corrections  (only the  comoving volume effect)  with a
         dotted line.  The symbol coding for the 15$\,\mu$m surveys is
         the same as in Fig.~\ref{figure:Diff_nomodel}.}
       \label{figure:Diff}
     \end{figure*} 
   %______________________________________________  
{Faint galaxy
counts  are computed  following the  already published  formalism (see
Sect. 2  in Guiderdoni \&  Rocca-Volmerange, 1990) which  assumes that
the number of  galaxies of each type is  conserved. Comparison of the
$ISO$-ESS  number  counts  with   the  predictions  of  P\'EGASE.3  at
12$\,\mu$m  is shown in  Fig.~\ref{figure:Int} for  cumulative counts,
and  in Fig.~\ref{figure:Diff}  for  Euclidean-normalized differential
counts  (solid  line in  both  graphs);  the  number counts  from  the
12$\,\mu$m $ISO$-ESS,  and from  the published 15$\,\mu$m  surveys are
also    shown   with    the    same   line/symbol    coding   as    in
Figs.~\ref{figure:Int_nomodel}  and  \ref{figure:Diff_nomodel}.   
Fig.~\ref{figure:Int}  shows that  the model  with ULIRGs  is  in good
agreement  with   the  \ISOc-ESS  cumulative   counts.   After  colour
correction, it also fits the deep \ISO counts at 15$\,\mu$m as well as
the  ultra-deep survey  down to  0.05  mJy in  the cluster-lens  A2390
(Altieri et al. 1999; Lemonon et al.  1998), implying that there is no
significant number  density variation between the  field and clusters.
The  differential  counts (Fig.   \ref{figure:Diff}),  which are  more
constraining, remain  in reasonably good  agreement with the  data, in
particular  at faint  fluxes.  The Euclidean-normalized  differential
counts  predicted  with P\'EGASE.3  do  show  the  departure from  the
Euclidean cosmology (horizontal  line) observed at 0.3 mJy  in all the
data  samples.   This bump  is  not due  to  the  evolution of  bright
spirals, nor to normal early-type  galaxies, but only to the evolution
of the  third of elliptical galaxies  (9\% of all  galaxies) which are
dusty ultra-bright ellipticals.   Only from a few 0.1 to $\simeq$1 mJy, 
the slope of
the Marano  Deep Field is respected  by models, in  excess relative to
other observations by a factor  2.  The model prediction with only the
universe  expansion  effect  (obtained  by applying  only  the  $k(z)$
corrections to the SEDs)  is also plotted in Fig.~\ref{figure:Int} and
Fig.~\ref{figure:Diff} as a dashed line: it is noticeably insufficient
to reproduce  the marked excess  counts and the  peak at 0.3  mJy.  We
also checked that  the model without any correction  (i.e.  no $k(z) +
e(z)$  corrections  applied to  the  SEDs),  which  only includes  the
evolution  of   the  comoving  elemental   volume,  yields  decreasing
differential  counts which  are incompatible  with  observations; this
curve is shown  in Figs.~\ref{figure:Int} and Fig.~\ref{figure:Diff} as
a dotted line.}
 
Note that  the comparison  of the observed  \ISOc-ESS counts  with the
Euclidean case is  more meaningful in the flux  range where the number
density  is the  highest. When  galaxy numbers  are  statistically too
small, error bars are large as  shown at fluxes higher than 10 mJy and
$\leq$ 0.1 mJy.   Finally we can ask whether  the normal spirals which
are  ultra-luminous in the  MIR ($\log  (L_*(12\,\mu m)/L_\odot)=9.8$\
and  $\log (L_*(24  \,\mu  m)/L_\odot)=9.6$\ (see  Table  1) can  also
reproduce the  bump of MIR  counts as well  as models with  ULIRGs. At
12$\,\mu$m, the  normal populations, dominated  in the MIR  by spirals
Sbc are quite unable to reproduce the excess counts of the MIR
surveys (Figs ~\ref{figure:Int}  and ~\ref{figure:Diff}, left), even by adding
9\% to the $\sim$57\% of normal  spirals in place of gas-poor elliptical galaxies. 

  %%%%%%%%%%%%%%%%%%%%%%%%%%%%%%%%%%%%%%%%%%%%%%%%%%%%%%%%%%%
 
  \section{Results of modeling Spitzer galaxy counts at 24 $\mu$m}  

  %==========================================================
  \subsection {The $Spitzer$\ surveys}

   %______________________________________________ 
     \begin{figure*}
 \includegraphics[width=7.5cm]{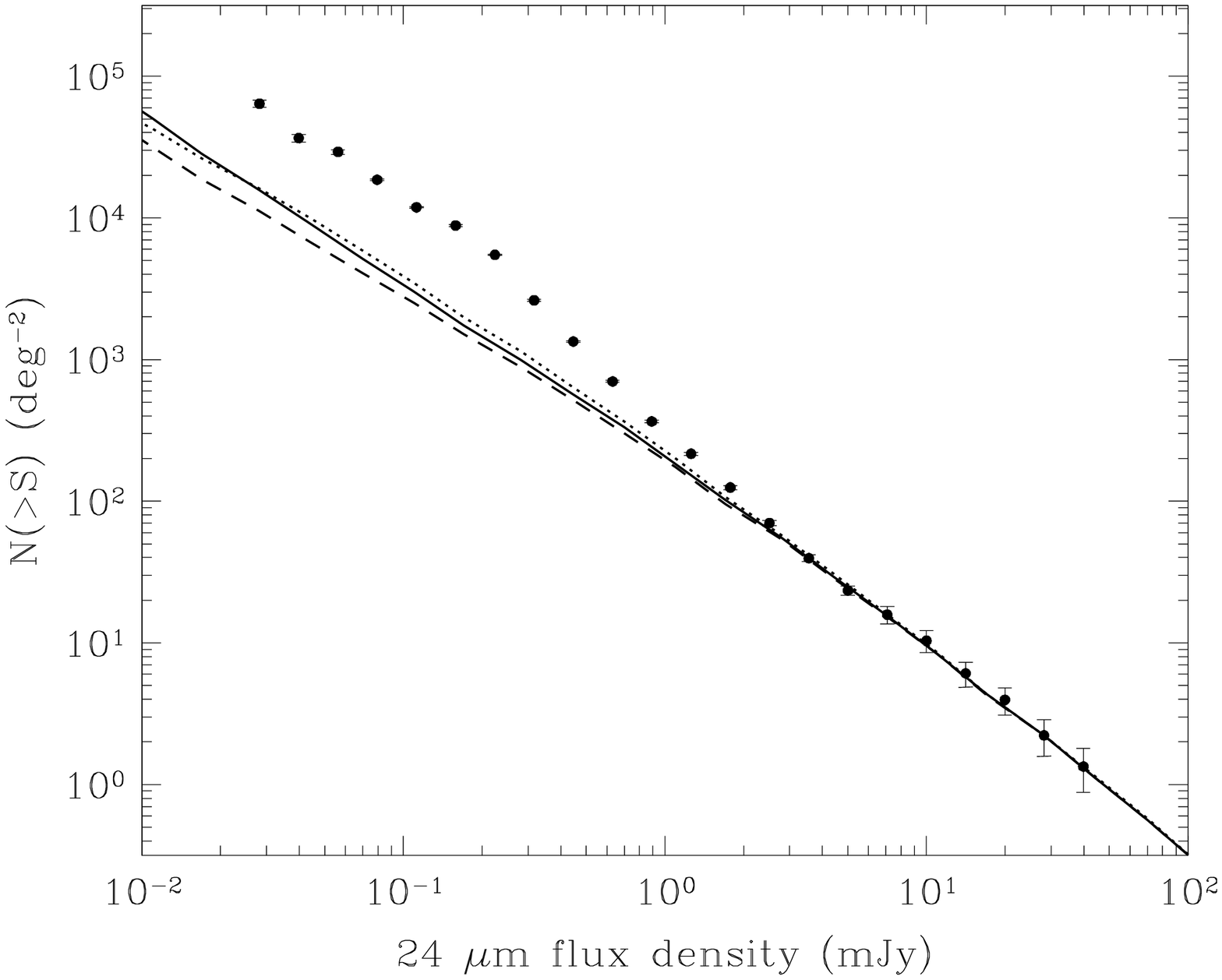}
\includegraphics[width=7.5cm]{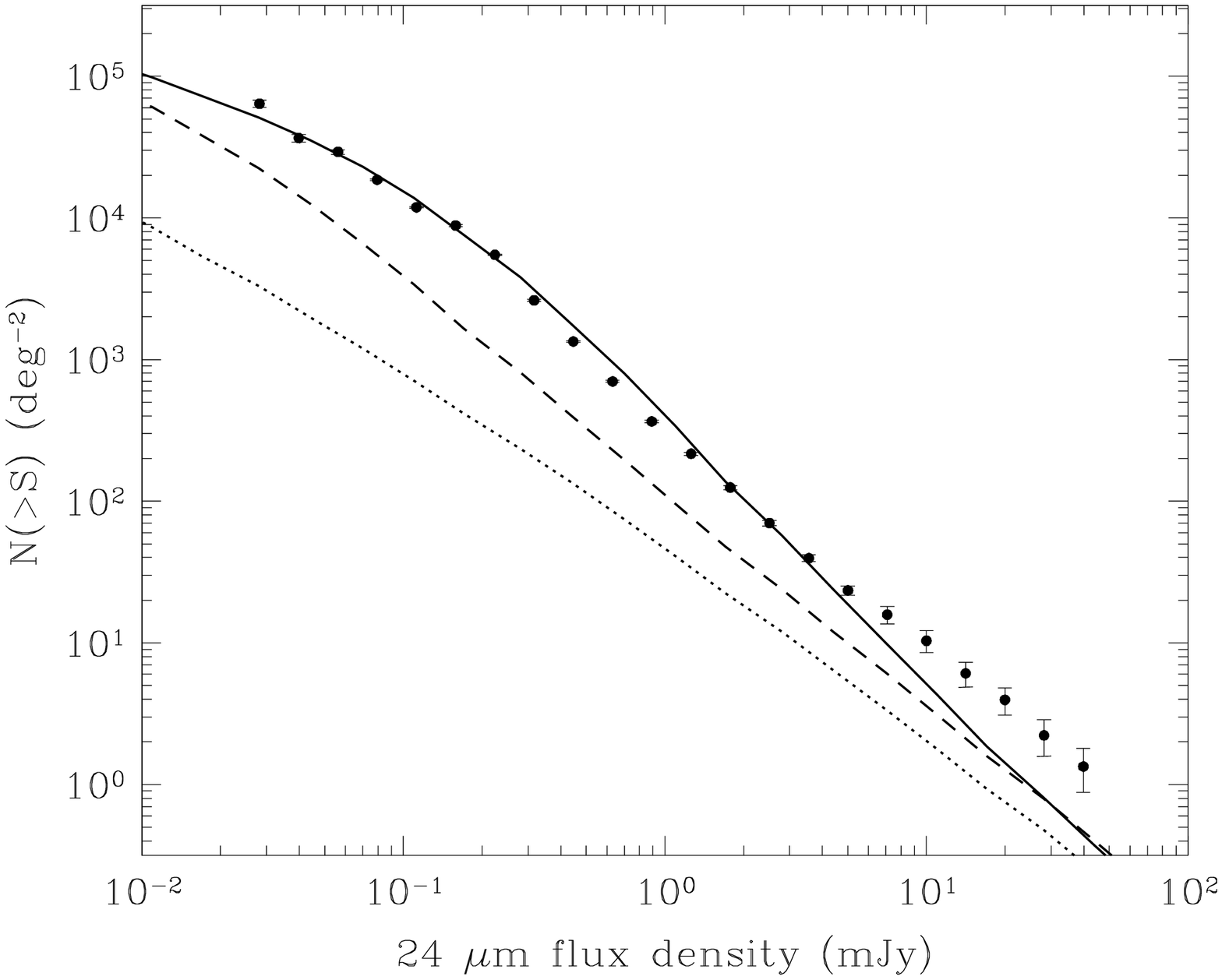}
     \caption{Cumulative  24 $\mu$m faint  galaxy  counts  (solid line) 
predicted  by model 1 (normal, left) and model 2 (with  ULIRGs, right) using 
P\'EGASE.3 (the  same models as  for the interpretation of the
$ISO$ 12 $\mu$m counts), are compared to the \Spitzerc/MIPS/24 $\mu$m 
observations by Papovich et al.
(2004).   The  luminosity function  is  from  $IRAS$/25 $\mu$m (Shupe  et
al.  1998).  As  in previous  figures,  the dashed  line is for counts taking 
into account only the $k$- correction  and volume expansion
effect,  while the  dotted  line corresponds  to  the comoving  volume
correction only (no $k+e$ correction).  }
     \label{figure:counts24mu}
\end{figure*}

   %______________________________________________
     \begin{figure*}
\includegraphics[width=7.5cm]{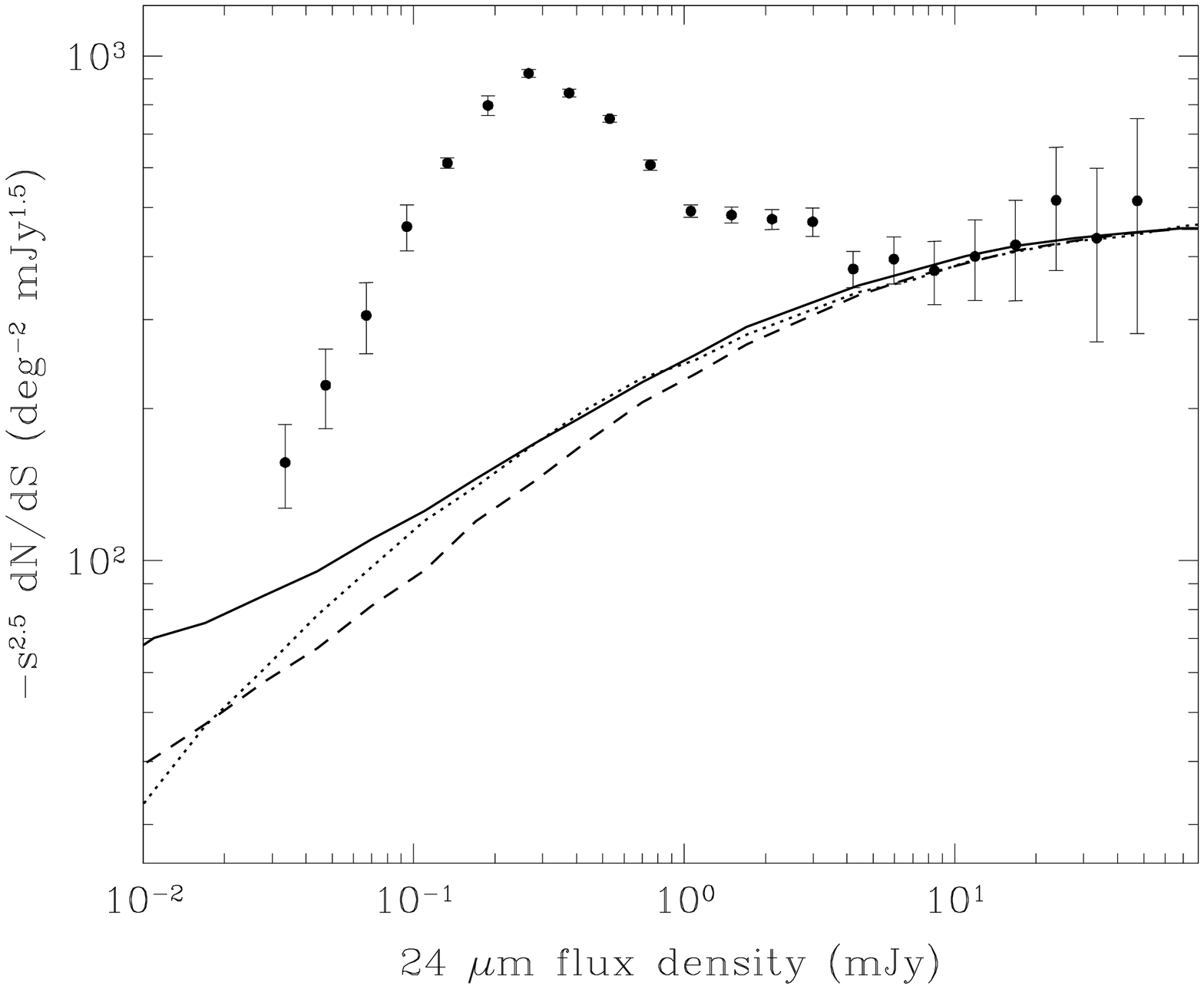}
\includegraphics[width=7.5cm]{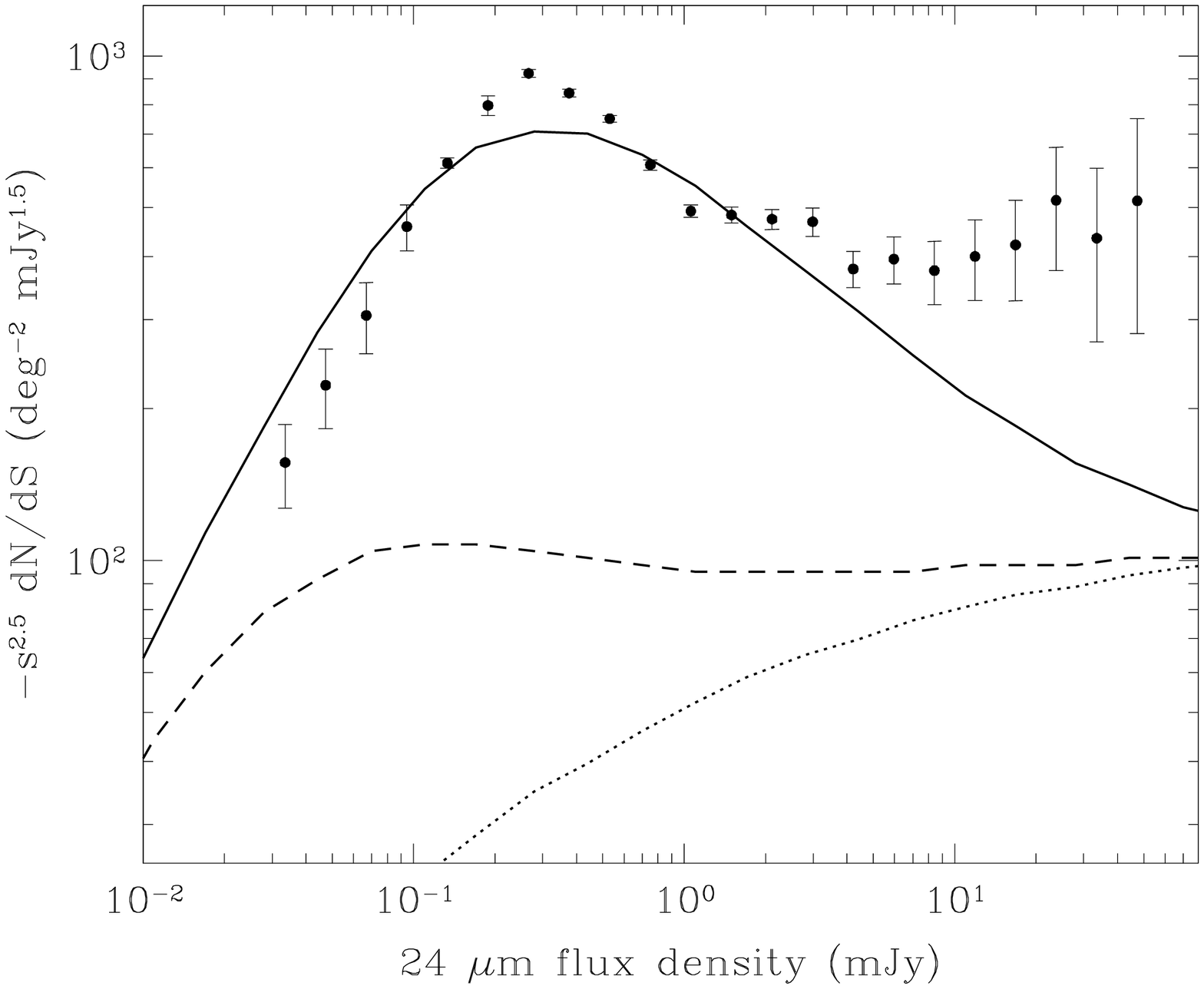} 
      \caption{The differential 24 $\mu$m faint galaxy counts computed 
          with P\'EGASE.3 (solid line) for model 1 with normal galaxies only
(left) and model 2 with  ULIRGs (right), both compared to  
          the $SPITZER$/MIPS/24 $\mu$m observations of Papovich et al. 
          (2004). The other symbols  are identical to the previous figure. 
For the 24 $\mu$m filter, the k-correction line (dashed-line, left) 
appears below the comoving volume line because the k-correction is negative, as shown 
by the SEDs displayed in Fig. \ref{figure:spirals}. } 
       \label{figure:counts24mudiff}
     \end{figure*}
   %______________________________________________ 

The deepest  {\it Spitzer}/MIPS  24 $\mu$m surveys are  in the  area of
the Chandra  Deep Field South  (Papovich et  al. 2004).  The corresponding
galaxy counts  (cumulative, shown in  Fig.~\ref{figure:counts24mu}, and
differential normalized    to   the   Euclidean case,   shown   in
Fig.~\ref{figure:counts24mudiff})     are    characterised     by a  typical bump  of  the galaxy
density between 3  and 0.03 mJy, similar to that  observed in the 12
$\mu$m  and 15  $\mu$m counts.   This  evolution signature  of the  24
$\mu$m  counts is  confirmed by  Marleau et  al. (2004)  and  Chary et
al. (2004). More recently,  in the GOODS-ELAIS-N1 field, Rodighiero et
al.   (2006) have lowered the  confusion  limit  by  about 30--50\%  using  a
deblending  technique,  which  leads  to  a  decrease  of  the  bright
differential counts by a factor  $\sim3$, and an increase of the count
slope at  faint fluxes.  The  statistics of the  24 $\mu$m observations
are  poor for  flux  densities $>10$  mJy  and the  samples suffer  of
incompleteness for fluxes $<80\ \mu$Jy.

  %==========================================================
  \subsection {The local IRAS/25 $\mu$m luminosity function} 

To model the  $Spitzer$/24~$\mu$m  counts,  we   use  the  
$IRAS$/25~$\mu$m luminosity function (corrected for 24 $\mu$m) 
as measured by Shupe et al. (1998),
\begin{equation}
\phi_{0}(L)=C\left(\frac{\alpha}{x}   +  \frac{\beta}{1   +  x}\right)
x^{1-\alpha}\left(1+x\right)^{-\beta},
\label{eq:lf2}
\end{equation} 
where  $x   =L/L_*$,  and   the  parameters  are   $\alpha=0.437$  and
 $\beta=1.749$ for all galaxy types. 
As for 12 $\mu$m, we compute
$L_*$(24 $\mu$m) by types from the Heyl et 
al. (1997) $L_*(B_J)$ and the  $B_J$-24$\mu$m colour
computed with P\'EGASE.3 (including filter correction from 
$IRAS$/25 $\mu$m to $Spitzer$/24 $\mu$m). We assign to spirals 
Sbc the value of L$_*$(24 $\mu$m) derived from Shupe et  al. (1998).
The same offset as for Sbc is applied to all types. 
 The resulting values  of 
 the characteristic $L_*$\ and the corresponding fractions per
 galaxy type are listed  in Table 1. 

  %==========================================================
  \subsection {Cumulative and differential counts with P\'EGASE.3 at $24 \mu$m}

We  model the  faint  galaxy counts  through  the $Spitzer$/24 $\mu$m
filter  with  the  code   P\'EGASE.3  and  the  same  evolving  galaxy
population (evolutionary scenarios, density fractions) as already used to
predict the 12  $\mu$m counts.  Fig.       \ref{figure:counts24mu} 
and Fig.\ref{figure:counts24mudiff} present the  comparison of the models 1 and 2
with  the  cumulative and  differential  number counts,  respectively,
obtained by Papovich et al. (2004). As for the 12 $\mu$m counts,
the 24 $\mu$m cumulative counts are well  reproduced from the faintest flux 
up   to  a   few   mJy.
Fig.~\ref{figure:counts24mudiff} shows that  the  marked  steepening of 
the  differential counts normalized  to Euclidean, and the subsequent 
decrease at faint
fluxes are predicted by  P\'EGASE.3, with a  peak at $\sim0.3$  mJy as
observed.  Note that the departure  of the model from the observations
at  bright  and  faint fluxes  is  in  good  agreement with  the  data
corrected  for incompleteness  and deblending  (see Rodighiero  et al.
2006). 
Once again, the fit is more meaningful in the flux range where the number
density is the  highest, as objects observed at  high fluxes in the
survey area are rare. Moreover, for the 24 $\mu$m filter, the $k$-correction
is negative at $z<2$ as shown in Fig. \ref{figure:spirals} by the 
SED slope from 24 $\mu$m to $\simeq$\ 8 $\mu$m.
In contrast  with the study by  Gruppioni et al. (2005),  based on the
flux density  ratio $S_{24\mu m}/S_{15\mu  m}$, our modeling  does not
require  a  population  of  additional starbursts, but rather the 
very strong evolution factor at high redshift of the star 
formation rate of elliptical galaxies as presented below. 

%%%%%%%%%%%%%%%%%%%%%%%%%%%%%%%%%%%%%%%%%%%%%%%%%%%%%%%%%%%%%%%%%%%%%%%%%%%
  \section{The cosmic star formation rate SFR(z) global per type }
   
%______________________________________________ 
     \begin{figure*}
\includegraphics[width=18cm]{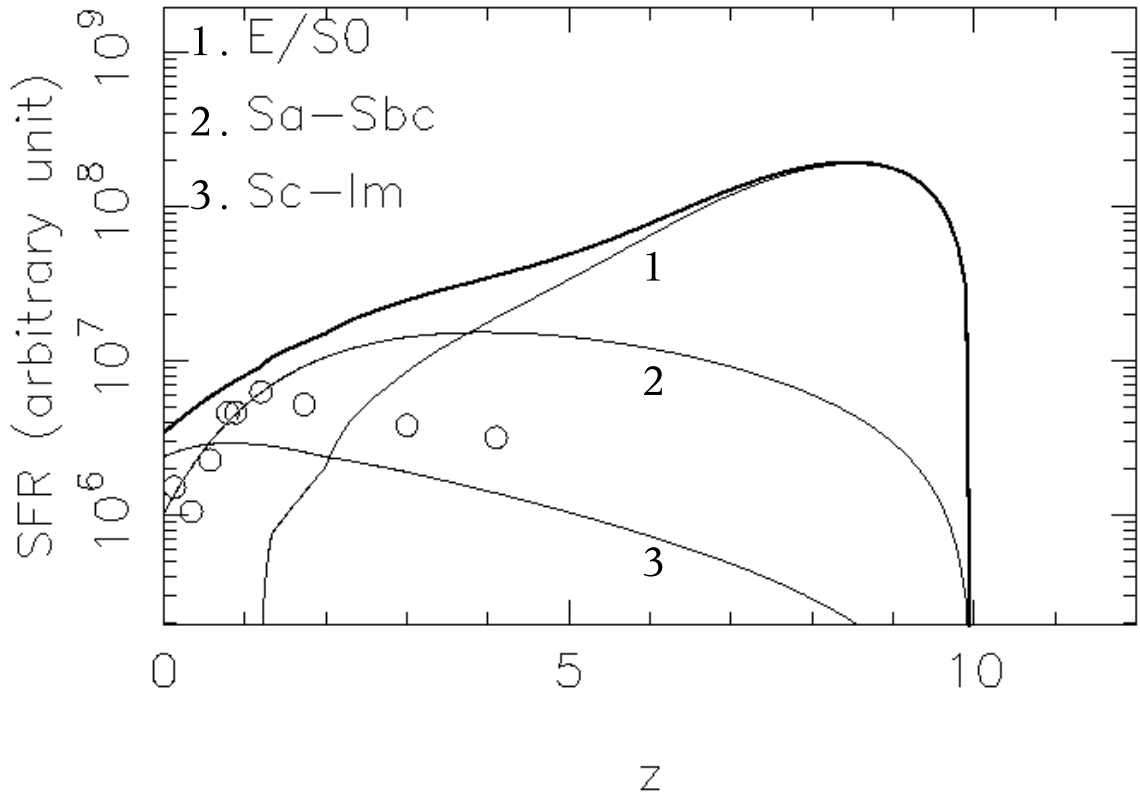}
         \caption{Histories of the  cosmic star-formation rate density
(global and by types) derived  from models of galaxy populations which
fit optical and IR faint galaxy counts.  The total SFR(z) (heavy solid
line),  summed on  all types  is shown.   The three  $SFR_i(z)$\ (thin
lines) correspond  to: 1)  the major SFR  at high reshifts  ($z>$4) is
from E/SO galaxies,  due to the bulk of stars  formed at early epochs,
its  contribution  at $z\simeq$0  is  null  because previous  galactic
winds.   2)  the intermediate  SFR  is  from  early spirals  (Sa-Sbc),
dominant from $z=0.6$ to $4$; it also fits the slope of the I-selected
CFRS observations  (empty circles), up to  $z$=1 before incompletness.
3) the weakest SFR are from  Sc-Im galaxy types (late-type spirals and
irregulars) except at $z\simeq$0  where they are numerous and actively
forming stars.   Models of ULIRGs follow the  scenario of ellipticals,
strongly evolving from $z>$2 to $z_{for}$ (here arbitrarily adopted at
10).   This  is   the   most  conservative   hypothesis  for   ULIRGs:
ultra-brightness is  likely attributed to active  nucleus while adding
occasional starbursts would increase the global $SFR(z)$.}
     \label{figure:SFR}
     \end{figure*} 
   %______________________________________________ 

The models  of galaxy populations which simultaneously fit the  
12 $\mu$m and 24
$\mu$m galaxy counts  can be used to predict the cosmic star 
formation  rate at high $z$. We then compute the global star formation rate
$SFR(z)$\ as,
   \begin{equation}
     SFR(z) = \Sigma_i\ SFR_i(z),
     \label{eq:SFH} 
   \end{equation}
          where $SFR_i(z)$  is the star formation  rate $SFR_i(z)$ per
galaxy type  $i$. The  total cosmic $SFR(z)$  (heavy line) and  SFR by
groups  of  galaxy  type  $SFR_i$(z)  (thin lines)  are  presented  in
Fig.  \ref{figure:SFR} in  arbitray units  (mass per  year  per volume
unit).  We distinguish the group 1. of E/SO galaxies which 
shows a striking SFR increase from
$z>1$ up  to the  redshift of formation  ($z_{for}$=10).  The  bulk of
stars formed at early epochs ($z>4$) while the SFR at $z\sim0$ is null
because  previous galactic  winds have  ejected all  the  gas content.
Then the  group 2. of  early  spirals
(Sa-Sbc) which is  dominant from $z=0.6$ to  $z\sim 4$ with  a predicted SFR
which increases from $z=0$ to 1 by a factor of $\sim10$, as found 
by Lilly et  al. (1996), Madau et al. (1996),  Connolly et al. (1997)
(empty circles), mainly from the $I$--band selected CFRS observations.
Finally the group 3. of late-type spirals and
irregulars (Sc-Im) is dominant from  $z\sim0$ to $z\simeq0.5$  but has
the  faintest  SFR at  $z>2$.  

Because ULIRGs  follow  the SFR  scenario  of ellipticals,  both
normal ellipticals and ULIRGs are  represented by the same line (E/S0)
on Fig.  \ref{figure:SFR}. This hypothesis, fully justified  if the origin
of the huge luminosity of ULIRGs is  not stellar (i.e. if it is due to
an  active nucleus), emphasizes  that the  star  formation history
shown in  Fig.~\ref{figure:SFR} is a  lower limit because it  does not
include   the  possible   contribution  from   starbursts   caused  by
interactions. Our modeling could  accept some such occasional starbursts as
long as they represent only a  few percents of the total stellar mass 
and are short.

The evolution of the SFR is different in  the two  groups Sa-Sbc  and
 Sc-Im. Between $z\simeq1.5$ and $z=0$, which corresponds 
to ages of $\simeq$ 6 to 13 Gyr,  the star  formation  rate of  a 
10$^{11}$ M$_{\odot}$  Sa
 spiral decreases from $\simeq$ 20 to 2 M$_{\odot}$ yr$^{-1}$ at $z=0$.
 In  the same redshift  interval, the  star formation  rate of  a 
less-evolved Sc spiral with the same mass increases from $\simeq 3$\ to 4.5
 M$_{\odot}$ yr$^{-1}$. This is consistent with the local observations:
 at   $z=0$,  the   current  mean  
 SFR  of   galaxies   is  $\simeq 5$\ $M_{\odot}$ yr$^{-1}$  for late  
spirals of  10$^{11}$ M$_{\odot}$ (and
 $\simeq 0.5$\ $M_{\odot} yr^{-1}$ for 10$^{10}$ M$_{\odot}$ irregulars);
 it is as low as 2 $M_{\odot}$ yr$^{-1}$ for early 10$^{11}$ M$_{\odot}$
 spirals (Kennicutt 1983).  This  SFR(t) variation is explained by the
 corresponding $M_{\text{gas}}$(t)  variation.  Because early  spirals formed
 stars  more efficiently  in  the past,  their  gas reservoir becomes  
 depleted  at $z\sim1.5$  and the  gas-dependent  SFR is  then rapidly
 decreasing by lack of fuel. This effect can  explain the apparent
 ``down-sizing'' of galaxies (Panter et al. 2006).

	 When summing over all galaxy types, the total cosmic $SFR(z)$
         (thick  line) increases  at high  redshift: it  evolves  by a
         factor  $\leq 3$  between $z=0$  and  1.  This  result is  in
         agreement with the gradual  decline of the UV luminosity from
         the  deep  multi-wavelength Hawaiian  surveys  (Cowie et  al.
         1999). This evolution is much smaller than  the factor 10 
         decrease of the SFR  derived from
         the  less deep  $I$-band selected CFRS  sample (Lilly  et  al. 1996;
         Tresse et al. 2002). The CFRS is an $I$-band selected sample,
         thus biased towards early  spirals, the major emitters in the
         $I$  band.  This  is  confirmed  by the  good  match between the
         $SFR(z)$\ of the CFRS (empty
         circles) and that  modelled  for the  Sa-Sbc  galaxies  in
         Fig.  \ref{figure:SFR}.   We  conclude that the $I$-band 
         selection  bias
         excludes late-type  galaxies, too  blue for detection  in the
         $I$-band at the depth of CFRS.
	 	
	  %%%%%%%%%%%%%%
%%%%%%%%%%%%%%%%%%%%%%%%%%%%%%%%%%%%%%%%%%%%
  \section{Discussion}

The faint galaxy counts at 12 $\mu$m in the \ISOc-ESS survey area show
the same typical signature of evolution at 0.3 mJy as already found in
the  \ISOc/15  $\mu$m  and  \Spitzerc/24 $\mu$m  galaxy  counts.   The
careful  flux calibration  of the  galaxy  catalogue is  based on  the
optical-MIR   statistical  properties  of   stars.  Our   results  are
consistent with the other MIR  counts at 15 $\mu$m, thus demonstrating
that  the surveyed  area of  ~800 arcminutes  square is  sufficient to
average the  inhomogeneities and  properly analyze the  populations of
galaxies.

The  most important result  of our analysis  is that a  minor ($<
10\%$)  population  of  dusty  ultra-bright  elliptical  galaxies  can
explain the  excess of the mid-IR  emission observed in  the 12 $\mu$m
and 24  $\mu$m faint galaxy counts  at $\simeq 0.3$ mJy.  Here, due to
its high IR brightness, this population is associated to ULIRGs, while
the other  populations, seen in the  UV-optical-nearIR as well  as in MIR
counts,  are called  normal  galaxies. Because  the evolutionary  code
PEGASE.3  predicts multi-wavelength SEDs by simultaneously following the  evolution of
stars,  gas, dust and  metal-enrichment, our  analysis results  find a
natural explanation in the basic scenarios of galaxy evolution.

The strong advantage of our analysis is the multi-wavelength approach:
the same SFR scenarios already  found to fit the UV-optical-nearIR galaxy
counts (Fioc \&  Rocca-Volmerange, 1999a) are also applied  to the MIR
(12 $\mu$m,  15 $\mu$m and 24  $\mu$m) using \ISO  (see companion article
and Elbaz et al. 1999) and \Spitzer satellites (Papovich et al.  2005;
Le  Floc'h et  al.  2005).   These evolutionary  scenarios  are robust
because the evolution time-scales  of the dominant emitters at various
wavelengths  (massive  stars,  evolved  stars, dust  grains  from  the
interstellar medium  and HII regions) go  from a few  million years to
$\simeq  13$ Gyr.   In the  mid-infrared, the  new model  with ULIRGs,
proposed  to fit $ISO$/12 $\mu$m  galaxy counts does not likely change the
UV-optical-nearIR  predictions; moreover  it  is confirmed  by the 
$ISO$/15$\mu$m and the more recent $Spitzer$/24 $\mu$m galaxy counts.

One  difficulty of  the interpretation  is that,  at $z\simeq  0$, the
brightest Sbc  spirals appear  as bright as  dusty ellipticals  in the
MIR.  Fig. \ref{figure:Int}  to Fig.  \ref{figure:counts24mudiff} show
that the populations of ``normal'' ellipticals, spirals and irregulars
(model 1)  seen in the  optical are largely insufficient  to reproduce
the 12 $\mu$m and 24  $\mu$m differential and cumulative counts.  Even
the                brightest               spirals               (with
$\log(L_*(12\,\mu\mathrm{m})/L_{\odot})=9.6$),     which    have    IR
luminosities  comparable to ULIRGs  at $z=0$,  decline too  rapidly at
increasing $z$ to fit the 12 $\mu$m and 24 $\mu$m number counts.

The surprising  result is that  we succeed in reproducing  the typical
excess of MIR counts observed at $\simeq0.3$ mJy by only replacing 9\%
of ``normal''  galaxies in  the optical (and  1/3 of  the ellipticals)
with ultra-bright  galaxies in the IR.  In fact at  high redshift, the
evolution  correction  takes  over  as  the  major  parameter:  it  is
noticeably insufficient for  spirals; only ellipticals have sufficient
star formation  rates to  reproduce the stellar  and dust  emission at
high  redshifts.  The fraction  of  these  objects  is small  and  the
enormous  luminosities required can  only be  reached if  these ULIRGs
contain huge dust masses heated by large numbers of energetic photons.

The  star  formation history  of  ULIRGs  follows  that of  elliptical
 galaxies in Fig.~\ref{figure:SFR}: large masses of stars and dust are
 formed  at  high  $z$. ULIRGs would  appear  as  ``normal''
 ellipticals  in  the  optical   with  masses  of  $\simeq$  10$^{12}$
 M$_{\odot}$; they could even be more massive if the excess of stellar
 luminosity is hidden by an excess of dust.

This  dusty   massive  population,  revealed  at  high   $z$\  in  the
mid-infrared,  evokes  the   population  of  high-$z$  ($>4$)  massive
ellipticals found in the  $K$-$z$\ Hubble diagram (Rocca-Volmerange et
al.   2004), and  confirmed in  the rest-frame  $H$-$z$\  diagram with
Spitzer  (Seymour  et  al.,  submitted): these  distant  galaxies  are
forming  also  high  masses of  stars  and  dust.   Due to  the  short
time-scales required to build such objects at $z > 4$ ($<$ 1Gyr), both
populations  likely  formed  at  early  epochs  by  rapid  dissipative
collapse or major merging,  rather than by slower hierarchical merging
which would take $\simeq$ 10 Gyr.

This interpretation of ULIRGs  as dusty massive  ellipticals is in agreement
with the drastic evolution  of the infrared  luminosity function
when compared  to the UV  luminosity function (Takeuchi et  al. 2006).
It is also in agreement with the result that the MIR-selected galaxies
contribute to more than 70\% of the Cosmic Extragalactic Background in
the MIR (Dole et al.  2006): our modelling confirms that other
galaxy types are too faint  at both 12$\,\mu$m and 24$\,\mu$m. However,
the conclusion of the last authors 
that galaxies contributing  the most to  the total
cosmic infrared  background have intermediate stellar  masses
is not confirmed by our results.

 One puzzling  issue is to know how the dust mass could be maintained
inside the  galaxy host, hence  no released by galactic winds
according in normal ellipticals.
   To maintain the large dust  mass of ULIRGs/dusty
ellipticals within galaxy, dust must  not be mixed with gas and stars.  It
has to  be located in preferential  zones such as the  galaxy core. In
the core, where  the potential well becomes intense in  case of an embedded
black  hole,  dust  could  be  preferentially retained.   In  the  AGN
environment, the deep potential well would drag dust more efficiently,
as it is more massive than gas; dust then would fall more rapidly down
in the inner core. Moreover if a Compton thick  AGN  is embedded within  the  proposed
ULIRG/dusty  ellipticals, the large  
variability  from  one ULIRG  to
another (Armus et al.  2006) is explained by orientation effects.
Only observations at high spatial
resolution will allow the dust geometry to be determined.

The  other issue  is the  presence of a large  number  of energetic
photons heating  dust grains at  all ages. To produce them, massive stars and/or
the presence of an AGN can be evoked. Our
results  do  not  exclude that,  in  the  case  of  galaxy
interactions, an  exceptionally extincted starburst,  undetected in the
optical, could  be ultra-bright in the  IR. But such an  event is rare
and is not representative of a galaxy population on a long time-scale. 
Note that  despite the high value
of their MIR
luminosity  ($\log(L_*(12\,\mu\mathrm{m})/L_{\odot})=9.9$)  at  $z=0$,
the  proposed ULIRGs  faintly contribute to the faint  UV-optical-NIR
galaxy counts by their number. They also are suffering an exceptional
extinction due to their dust amount. Further spectral syntheses and 
high spatial resolution are clearly needed.
 The recent analysis  by Takeuchi et al. (2006), based on
the combination  of data  from the UV  satellite {\it GALEX}  and from
$IRAS$\ shows that  the luminosity function evolves more  strongly in the
far-infrared than  in the  far-UV. This is  compatible with  our dusty
elliptical population.  At last, by analyzing the far-UV galaxy counts
with FOCA at 2000\AA, Fioc \& Rocca-Volmerange (1999a) suggested that a
fraction of episodic starbursts could  be required to interpret the UV
excess of  galaxy counts, in addition to the normal  populations of
galaxies.  However, the number  density, weak star formation rate, and
low metallicity of these populations are not sufficient to explain the
excess of MIR luminosity at high redshifts.

Finally, these scenarios of ultra-luminous galaxies at high redshift,
imply a very  rapid phase of mass accumulation.  This is also supported
by the fact that ULIRGs,  evolving as ellipticals and hosting a hidden
AGN, look like the  population  of  radio-galaxies
discovered from the  $K$-$z$ diagram at high redshifts which also show strong
and  hot  dust  signatures  (Rocca-Volmerange \& Remazeilles  2005).
However, the proposed population of ULIRGs derived from the infrared is
more  numerous (9\%)  than  the  radio-galaxy  hosts  detected in  the
optical  ($<4$\%). This  indicates that  half of  the ULIRGs  could be
so obscured  in  the optical that they would be invisible.  They  may  
however  be revealed  at  other
wavelengths. Several surveys  have discovered populations of AGN which appear 
brighter and  denser than the classical populations  identified in the
optical  (see for  example Martini  et al.  2006).  The  population of
hyper-LIRGs ($L_{IR} > 10^{13}L_\odot$), sometimes
associated  to ${\text{Ly}_{\alpha}}$\  blobs,
has a  low $L_{\text{Ly}_{\alpha}}/L_{\text{bol}}$\ efficiency  (0.05--0.2\%)
according to Colbert et al (2005).  The  12$\,\mu$m  and 24$\,\mu$m
galaxy  counts analysed  here may  correspond to  the  best wavelength
domain where this population  of  possibly  embedded
AGNs could be detected. 

%%%%%%%%%%%%%%%%%%%%%%%%%%%%%%%%%%%%%%%%%%%%%%%%%%%%%%%%%%%

  \section{Conclusion}
   
   We present the  faint galaxy  counts derived  at 12$\,\mu$m
from the observation of a  large and deep mid-infrared (MIR) survey in the
field of the optical ESO-Sculptor Survey (ESS), through the large LW10
filter with  the ISOCAM instrument  on board the \ISO  satellite. The
infrared  observations  cover  an area of $\sim75$\%  of  the  ESS  spectroscopic
survey,  where  galactic  cirrus is  sparse, and  were  performed  in
continuous  raster mode.  The  flux calibration  has been  adjusted by
using the  optical-infrared IRAS colours  ($B-12\,\mu$m, $V-12\,\mu$m)
of standard stars.  
Because of its
large area of  $\sim680$  arcmin$^2$,  the \ISOc-ESS  survey
provides complete  12$\,\mu$m galaxy counts  down to 0.24 mJy, after
incompleteness corrections  (using two independent  methods). The full
data analysis and  the resulting catalogue of 142  detected sources is
published in the companion paper, Seymour et al. (2007).
 
The  galaxy counts  are presented  using two  different  binnings: 
cumulative  counts $N(>S)$, to  avoid the  fluctuations in  the number
density  per bin,  and Euclidean  normalized differential  counts
-$S^{2.5}dN/dS$.    When  corrected   for  incompleteness,   both  the
cumulative and differential  \ISOc-ESS 12$\,\mu$m counts averaged over
the $\sim680$  arcmin$^2$ area show  good agreement with  the existing
measurements  in  the close-by  \ISOCAM  filter  at 15$\,\mu$m,  after
correction  for   the  different  wavelengths.    In  particular,  the
Euclidean-normalized  differential counts of  the \ISOc-ESS  survey
display the  same excess as the  other existing MIR  surveys at
flux densities of ~0.3 mJy.  This excess is also observed in the
$Spitzer$ galaxy counts at 24$\,\mu$m.

We propose an interpretation of the cumulative and differential
counts with the
help of the new evolutionary code P\'EGASE.3 (Fioc et al.  2007).  For
each galaxy type, P\'EGASE.3 predicts the 
spectral  energy distributions  from  the optical  to  the far-IR; 
the emission of stars and dust, the extinction, star formation
history, metal enrichment and dust mass are computed consistently. 

With these evolutionary standard  scenarios we have
successfully modelled the multi-wavelength  faint galaxy counts in the
far-UV, optical  and near-infrared (Fioc  \& Rocca-Volmerange 1999a).
In  the present article, we  are able to  simultaneaously fit the
\ISOc-ESS  12$\,\mu$m, $ISO$ 15$\,\mu$m  and the  \Spitzerc24 $\,\mu$m  faint
counts, by increasing  by $\simeq$  2.5 mag (at 12$\,\mu$m) to 5 mag
(at 24$\,\mu$m)  the luminosity  of a
small fraction  of galaxies (9\%;  all of elliptical type),  while the
rest of  galaxies (17.5\% normal ellipticals, 57\%  spirals and 16.5\%
irregulars) are identical to the galaxy populations already known from
the  UV-optical-NIR surveys.   The ultra-bright galaxies display all  the
characteristics of ULIRGs and appear as ``normal'' ellipticals in the
optical.  Because  these results cover a very  large wavelength domain,
from  the  UV-optical-NIR  to  the  MIR  (12$\,\mu$m,  15$\,\mu$m  and
24$\,\mu$m),  they confirm  the robustness  of our  scenarios. 

The  other important  point is  that no additional starbursts are 
required  to  fit  the  MIR excess.   Highly
luminous starbursts with a  short e-folding time ($\simeq 10^7$ years),
which may explain some nearby  ULIRGs such as M82, are not incompatible
with our  results if they  remain exceptional  objects.  Another
point is that  the normal galaxy populations, including  the bright IR
spirals, are insufficient at  high redshifts to fit  either the
cumulative  or  the  differential  \ISOc-ESS  12$\,\mu$m  and  SPITZER
24 $\,\mu$m counts.

The  star  formation  history   of  the  proposed  ULIRGs  and  normal
ellipticals, in respective proportions 1/3 and  2/3, can fully 
explain the excess
in  the cumulative and  differential galaxy  counts at  12$\,\mu$m and
24$\,\mu$m, thanks to a  large dust mass  produced by early-formed
stars. We suggest that in these  ULIRGs, dust is not mixed  with stars,
as  otherwise  it  would  be  expelled  by  the  galactic  winds.  The
possibility  of  an  embedded  Compton  thick AGN,  which  would explain the
ultra-brigthness by keeping dust within galaxy, will require
confirmation with other observations.   Similar to the  distant radio galaxies
found in the $K-z$\ diagram  (Rocca-Volmerange et al.  2004), the most
distant  ULIRGs  would appear  as  the  most  massive ellipticals.  As
concluded in the mentioned article,  this new result in the MIR favors
the hypothesis of a galaxy evolution process  based on dissipative collapse or a rapid merging
on a short time-scale ($<1$ Gyr) at high redshift ($z>5$).

We emphasize that these results are robust since they are derived from
the  simultaneous  adjustment of  the  MIR  12$\,\mu$m and  24$\,\mu$m
counts respecting the majority ($>90$ \%) of galaxies observed 
in the optical  faint number counts (down
to  $B>29$  from the  HDF-N,  William  et  al.  1996).  Moreover,  our
analysis by type allows us to identify the various factors explaining 
the steep  increase of the  faint galaxy counts.  The  comoving volume
element effect,  as well as  the expansion effect  ($k$-correction), are
insufficient  to explain the  faint count  excess.   Higher spectral and spatial
resolution observations associated to deeper  counts at  longer
wavelengths  with  the  $Spitzer$  satellite and  the  future  \Herschel
satellite  will hopefully  allow the  detection of  the  dust emission
emitted  by  early  elliptical  galaxies in  their  primitive  epochs,
validating  our present results  and allowing to hopefully observe 
primeval galaxy
populations down to the deepest extragalactic backgrounds.

%______________________________________________________________
   %\begin{enumerate}
   %\end{enumerate}
   
   \begin{acknowledgements}
N. Seymour acknowledges financial support from
the European Network `Probing the Origin of the Extragactic Background Radiation' (POE)
while at the Institut d'Astrophysique de Paris (IAP). 
%We wish to thank Michel Fioc at 
%IAP for facilitating the use of the code P\'EGASE.3 before publication. 
We also thank Prof. L. Woltjer for his stimulating interest 
for this \ISO observational program. 
   \end{acknowledgements}

   % Two column figure (Contours ISOCAM +optical)
   

\begin{thebibliography}{}
     
   \bibitem[1999]{altieri} Altieri, B., Metcalfe, L., Kneib, J.~P., McBreen, B., Aussel, H., 
     Biviano, A., Delaney, M., Elbaz, D., Leech, K., L\'emonon, L., and 3 coauthors 1999, A\&A, 343, 65
     
   \bibitem[1997]{arnouts} Arnouts, S., de Lapparent, V., Mathez, G., Mazure, A., Mellier, Y., 
     Bertin, E., Kruszewski, A. 1997, A\&AS, 124, 163
     
   \bibitem[1999]{aussel} Aussel, H., Cesarsky, C.~J., Elbaz, D., Starck, J.~L.
     1999, A\&A, 342, 313
        
   \bibitem[1995] {bellanger95a} Bellanger, C., de Lapparent, V., Arnouts, S., Mathez, G.,
     Mazure, A., Mellier, Y. 1995, A\&AS, 110, 159	
     
   \bibitem[1995] {bellanger95b} Bellanger, C., de Lapparent, V. 1995, ApJ, 455, L103
   
\bibitem[1996]  {bertin} Bertin, E., Arnouts, S., 1996, A\&AS, 117, 393
          
   \bibitem[1996]{cesarsky} Cesarsky, C.~J., Abergel, A., Agnese, P., 
     Altieri, B., Augueres, J.~L., Aussel, H., Biviano, A., Blommaert, J., 
     and 58 coauthors, 1996, A\&A, 315, 32
     
   \bibitem[2006]{caputi} Caputi, K., Dole, H., Lagache, G., McLure, R. J., 
   Puget, J. L., Rieke, G. H., Dunlop, J. S., Le Floc'h, E., Papovich, C., P\'erez-Gonzalez, P. G., 
   2006, ApJ, 637, 727

   \bibitem[2005]{chary2005} Chary, R.-R., Stern, D., Eisenhardt, P. 2005, ApJ, 635, L5

   \bibitem[2004]{chary2004} Chary, R.-R., Casertano, S., Dickinson, M. E., Ferguson, H. C., 
Eisenhardt, P. R. M., Elbaz, D., Grogin, N. A., Moustakas, L. A., Reach, W. T., Yan, H.S 2004, ApJ Sup. Ser., 154, 80
     
   \bibitem[1999]{clements} Clements, D. L., Desert, F.-X., Franceschini, A., Reach, W. T., 
     Baker, A. C., Davies, J. K., Cesarsky, C.  1999, A\&A, 346, 383
 
\bibitem[1997]{connolly} Connolly, A. J., Szalay, A. S., Dickinson, M., Subbarao, M. U., Brunner, R. J. 1997, ApJL, 486, 11

\bibitem[1980]{Couch} Couch, W. J., Newell, E. B. 1980, PASP, 92, 746
	
    \bibitem[1999]{cowie}  Cowie, L. L., Songaila, A., Barger, A. J. 1999, AJ, 118, 603
    \bibitem[2006]{dole}  Dole, H., Lagache, G., Puget, J.L.,  Caputi, K.I., 
     Fern\'andez-Conde, N.,  Le Floc'h, E., Papovich, C., P\'erez-Gonz\`alez, P.G., 
     Rieke, G.H., Blaylock, M. 2006, A\&A, 451, 417

    \bibitem[2003]{draine}  Draine, L., 2003, Annual Review Astron.Astrophys., 41, 241 

   \bibitem[2004]{egami} Egami, E., Dole, H., Huang, J.-S., P\'erez-Gonzalez, P., Le Floc'h, E., 
   Papovich, C., Barmby, P., Ivison, R. J. and 19 coauthors, 
   2004, ApJS, 154, 130

   \bibitem[1999]{elbaz} Elbaz, D., Cesarsky, C.~J., Fadda, D., Aussel, H., D\'esert, F.~X., 
     Franceschini, A., Flores, H., Harwit, M., Puget, J.~L., Starck, J.~L., and 4 coauthors 
     1999, A\&A, 351, L37
     
   \bibitem[1998]{fang} Fang, F., Shupe, D.~L., Xu, C., Hacking, P.~B. 1998, ApJ, 500, 693
     
  \bibitem[1997]{fioc97}  Fioc, M., Rocca-Volmerange, B., 1997, A\&A, 326, 950

   \bibitem[1999]{FRV99} Fioc, M., Rocca-Volmerange, B., 1999a, A\&A, 344, 393	
     
   \bibitem[1999]{fioc99} Fioc, M., Rocca-Volmerange, B., 1999b, astro-ph/9912179

  \bibitem[2007]{fioc07} Fioc, M., Rocca-Volmerange, B., Dwek, E., 2007, near to submission to A \& A
     
   \bibitem[1999]{flores} Flores, H., Hammer, F., Thuan, T.~X., Cesarsky, C., Desert, F.~X., 
Omont, A., Lilly, S.~J., Eales, S., Crampton, D., Le F\`evre, O. 1999, ApJ, 517, 148

%\bibitem[2001] {franceschini} Franceschini, A.; Aussel, H.; Cesarsky, C. J.; Elbaz, D.; Fadda, D., 
%	2001, A\&A, 378, 1

   \bibitem[1998]{galaz} Galaz, G., de Lapparent, V. 1998, A\&A, 332, 459
     
   \bibitem[2005] {gruppioni}  Gruppioni, C., Pozzi, F., Lari, C., Oliver, S., Rodighiero, G,
2005, ApJL, 618, 9

   \bibitem[1989] {Guhatha} Guhathakurta, P., Draine, B. T.  1989,
ApJ, 345, 230 

     \bibitem[2005] {GRV90}  Guiderdoni, B.; Rocca-Volmerange, B. 1990,
A\&A, 227, 362

\bibitem[1998]{hauser} Hauser, M. G., Arendt, R. G., Kelsall, T., Dwek, E., Odegard, N.,
     Weiland, J. L., Freudenreich, H. T., Reach, W. T. and 10 coauthors 1998, ApJ, 508, 25	
	
\bibitem[1997]{heyl} Heyl, J., Colless, M., Ellis, R. S., Broadhurst, T.
 1997, MNRAS, 285, 613

%% \bibitem[1999]{hogg99} Hogg, D.~W., 1999, astro-ph/9905116

    \bibitem[2002]{hu2002} Hu, E., M., Cowie, L. L., McMahon, R. G., Capak, P., Iwamuro, F.,
Kneib, J.-P., Maihara, T., Motohara, K. 2002, ApJL, 568, 75

   \bibitem[1983] {kennicutt} Kennicutt, R. C., Jr., 1983, ApJ, 272, 54

   \bibitem[1996]{kessler} Kessler, M.~F., Mueller, T.~G., Leech, K., Arviset, C., 
     Garcia-Lario, P., Metcalfe, L., Pollock, A., Prusti, T., Salama, A. 2003, ESASP, 1262
     
   \bibitem[2004]{lafranca} {La Franca}, F., {Gruppioni}, C., {Matute}, I., {Pozzi}, F., 
     {Lari}, C., {Mignoli}, M., {Zamorani}, G., {Alexander}, D.~M., {Cocchia}, F., 
     {Danese}, L., {Franceschini}, A., {H{\' e}raudeau}, P. and 7 coauthors  2004, \aj, 127, 3075 
     
   \bibitem[2004]{lapparent04} de Lapparent, V., Arnouts, S., Galaz, G., Bardelli, S., 
     2004, A\&A, 422, 841
     
   \bibitem[2003]{lapparent03} de Lapparent, V., Galaz, G., Bardelli, S., Arnouts, S., 
     2003, A\&A, 404, 831
     
   \bibitem[2006]{lapparent} de Lapparent, V., Seymour, N., Rocca-Volmerange, B. 2007, in preparation
     
   \bibitem[2004]{lapparent05} de Lapparent, V., Slezak, E. 2007, A\&A, submitted
     
   \bibitem[2002]{leborgne} Le Borgne, D., Rocca-Volmerange, B.,  2002,
     A\&A, 386, 446
     
   \bibitem[2005]{lefloch} Le Floc'h, E., Papovich, C., Dole, H., Bell. E. , Lagache, G., Rieke, G., 
     and 11 co-authors, 2005, ApJ, 632, 169
     
   \bibitem[1998]{lemonon} Lemonon, L., Pierre, M., Cesarsky, C.J., Elbaz, D., Pello, R., 
     Soucail, G., Vigroux, L., 1998, A \& A 334, L21
     
%%   \bibitem[1995]{lefevre} Le Fevre, O., Crampton, D., Lilly, S. J., Hammer, F.,
%%     Tresse, L.,  1995, \apj 455, 60


 \bibitem[1996]{lilly} Lilly, S. J., Le Fevre, O., Hammer, F., Crampton, D. 1996, ApJL, 460, 1

\bibitem[1996] {madau} Madau, P., Ferguson, H. C., Dickinson, M. E., Giavalisco, M., Steidel, C. C., Fruchter, A.  1996, MNRAS, 283, 1388

    \bibitem[2004]{Marleau}  Marleau, F. R., Fadda, D., Storrie-Lombardi, L. J., Helou, G., Makovoz, D., Frayer, D. T.,
Yan, L., Appleton, P. N. and 13 coauthors,  2004, ApJS, 154, 66	
         
   \bibitem[1997] {moneti} Moneti, A., Breitfellner, M. G., 1997,  Astrophysics \& Space 
     Science L. Ser., 210, 205 
     
   \bibitem[1997]{oliver97} Oliver, S.~J., Goldschmidt, P., Franceschini, A., Serjeant, S.~B.~G., 
     Efstathiou, A., Verma, A., Gruppioni, C. and 15 coauthors 1997, MNRAS, 289, 471O
     
   \bibitem[2000]{oliver00} Oliver, S., Rowan-Robinson, M., Alexander, D.~M., Almaini, O., 
     Balcells, M., Baker, A.~C., Barcons, X., Barden, M., Bellas-Velidis, I., Cabrera-Guerra, F.,
     and 60 coauthors 2000, MNRAS, 316, 749
     
   \bibitem[2002]{oliver02} Oliver, S., Mann, Robert G., Carballo, R., Franceschini, A., 
     Rowan-Robinson, M., Kontizas, M., Dapergolas, A., Kontizas, E. and 9 coauthors 2002, MNRAS, 332, 536
     
 \bibitem[2006]{panter} Panter, B., Jimenez, R., Heavens, A. F., Charlot, S. 2006, astro-ph/0608531

   \bibitem[2004]{papovich} Papovich, C., Dole, H., Egami, E., Le Floc'h, E., and 
     18 coauthors 2004, ApJS, 154, 70
     
   \bibitem[2005] {pearson} Pearson, C. 2005, MNRAS, 358, 1417
     
%   \bibitem[2005] {perezgonzalez} P\'erez-Gonz\'alez, P.~G., Rieke, G.~H., Egami, E., 
%     Alonso-Herrero, A., Dole, H., Papovich, C.,  Blaylock, M., Jones, J., Rieke, M., Rigby, J., 
%     Barmby, P., Fazio, G.~G., Huang, J.,  Martin, C., ApJ, in press (astro-ph/0505101)
     
   \bibitem[2003]{pozzi03} Pozzi, F., Ciliegi, P., Gruppioni, C., Lari, C., H\'eraudeau, P.,
     Mignoli, M., Zamorani, G., Calabrese, E., Oliver, S., Rowan-Robinson, M., 2003, MNRAS, 343, 1348
     
   \bibitem[2004]{pozzi04} {Pozzi}, F., {Gruppioni}, C., {Oliver}, S., {Matute}, I., 
     {La Franca}, F., {Lari}, C., {Zamorani}, G., {Serjeant}, S., {Franceschini}, A.,
     {Rowan-Robinson}, M. 2004, \apj, 609, 122
     
   \bibitem[2005]{press} Press, W. H., Schechter, P., 1974, ApJ, 187, 425

  \bibitem[1996]{puget96}  Puget, J.-L., Abergel, A., Bernard, J.-P., Boulanger, F., 
     Burton, W. B., Desert, F.-X., Hartmann, D., 1996, A\&A, 308, L5	
     
   \bibitem[1989]{puget89}  Puget, J.-L., Leger, A., 1989, ARA\&A, 27, 161
     
   \bibitem[2005]{rana} Rana, N.C., Basu, S., 1992, A\&A, 265, 499
     
        
   \bibitem[2005]{rocca5} Rocca-Volmerange, B., Remazeilles, M. 2005, 
A\&A, 433, 73

\bibitem[2004]{rocca4} Rocca-Volmerange, B., Le Borgne, D., De Breuck, 
C., Fioc, M., Moy, E., 2004, A\&A, 415, 931	
     
   \bibitem[1999]{rocca0}Rocca-Volmerange, B, 1999, in ``Toward a new millenium in galaxy morphology'',
     Block et al., ed. Kluwer, p. 238, reprinted from Astrophysics and Space Science, 1999, 
     vol 269-270, Nos 1-4.
     
   \bibitem[1988]{rocca8} Rocca-Volmerange, B., Guiderdoni, B., 1988. A\&AS, 75, 93

   \bibitem[2006]{Rodig2006} Rodighiero G., Lari C., Pozzi P., Gruppioni C., Fadda D., 
Franceschini A., Lonsdale C., Surace J., Shupe D., Fang, F., 2006, MNRAS, 371, 1891

    \bibitem[2004]{Rodig2004} Rodighiero, G., Lari, C., Fadda, D., Franceschini, A., 
     {Elbaz}, D., {Cesarsky}, C. 2004, \aap, 427, 773
     
   \bibitem[1999]{rrobinson99} Rowan-Robinson M., Oliver, S., Efstathiou, A., Gruppioni, C., 
     Serjeant, S., Cesarsky, C.~J., Danese, L., Franceschini, A., Genzel, R., Lawrence, A., 
     and 11 coauthors, 1999, in The Universe as Seen by \ISOc. Eds. P. Cox \& M. F. 
     Kessler. ESA-SP 427, p. 1011
     
   \bibitem[2004]{rrobinson04} Rowan-Robinson, M.; Lari, C.; Perez-Fournon, I.; Gonzalez-Solares, E. A.; La Franca, F.; Vaccari, M.; Oliver, S.; Gruppioni, C.; Ciliegi, P.; H\'eraudeau, P.; and 69 coauthors, 2004, MNRAS, 351, 1290
   
\bibitem[1993]{rush} Rush, B., Malkan, M.~A.,  Spinoglio, L. 1993, ApJS, 89, 1	
     
   \bibitem[2003]{sato} Sato, Y., Kawara, K., Cowie, L.~L., Taniguchi, Y., Sanders, D.~B., 
     Matsuhara, H., Okuda, H., Wakamatsu, K., Sofue, Y., Joseph, R.~D., Matsumoto, T.
     2003, A\&A, 405, 833
          
   \bibitem[2000]{serjeant} Serjeant, S., Oliver, S., Rowan-Robinson, M., Crockett, H., 
     Missoulis, V., Sumner, T., Gruppioni, C., Mann, R.~G. and 15 coauthors 2000, MNRAS, 317, 29
     
   \bibitem[2006]{seymour} Seymour, N., Rocca-Volmerange, B., de Lapparent, V. 2007, 
     (companion article), submitted
     
    \bibitem[1998]{shupe} Shupe, D.L., Fang, F.,  Hacking, P.B., Huchra, J.~P. 1998,
     ApJ, 501, 597

  \bibitem[1998]{silva} Silva, L., Granato, G.L., Bressan, A., Danese, L.,
1998, ApJ, 509, 103 

\bibitem[1984]{soifer} Soifer, B.T., Helou, G., Lonsdale, C., Neugebauer, G., Hacking, P., 
     Houck, J.R., Low, F.J., Rice, W., Rowan-Robinson, M., 1984, ApJL, 283, 1  
     
   \bibitem[2004]{somerville} {Somerville}, R.~S., {Lee}, K., {Ferguson}, H.~C., 
     {Gardner}, J.~P., {Moustakas}, L.~A., {Giavalisco}, M. 2004, ApJL, 600, 171
     
   \bibitem[2003]{spergel03} {Spergel}, D.~N., {Verde}, L., {Peiris}, H.~V., {Komatsu}, E., 
     {Nolta}, M.~R., {Bennett}, C.~L., {Halpern}, M., {Hinshaw}, G., 
     {Jarosik}, N. and 8 coauthors 2003, ApJS, 148, 175
     
   \bibitem[1998]{starck} Starck, J.~L., Abergel, A., Aussel, H., Sauvage, M., Gastaud, R., 
     Claret, A., Desert, X., Delattre, C., Pantin, E. 1999, A\&AS, 134, 135
     
   \bibitem[2006]{takeuchi} Takeuchi, T. T., Buat, V., Burgarella, D., 
astro-ph/0611796

\bibitem[1997]{taniguchi}Taniguchi, Y., Cowie, L.~L., Sato, Y., Sanders, D.~B., Kawara, K., 
     Joseph, R., Okuda, H., Wynn-Williams, C.~G. and 6 coauthors 1997, A\&A, 328,9
     
   \bibitem[2002]  {Tresse} Tresse, L., Maddox, S.J., Le F\`evre, O., Cuby, J.G. 
     2002, MNRAS, 337, 369
     \bibitem[2000]  {verstrate00} Verstraete, L., Pech, C., Moutou, C., Wright, C. M., Drapatz, S., L\'eger, A., 2000, ``The 2nd \ISO workshop on analytical spectroscopy''.Eds. A. Salama, M.F.Kessler, K. Leech \& B. Schulz. ESA-SP 456. p.319
%%   \bibitem[1997]{waters} Waters, L.~B.~F.~M., Cote, J., Aumann, H.~H. 1987, A\&A, 172, 225
   \bibitem[1997]{werner} Werner, M.W., Roellig, T. L., Low, F. J., Rieke, G. H., Rieke, M., 
Hoffmann, W. F., Young, E., Houck and 18 coauthors
2004, ApJS, 154, 1
    
   \bibitem[1972]{weinberg} Weinberg, S., 1972, ``Gravitation and cosmology: Principles and 
     applications of the general theory of relativity'' (New York: Wileys)
     
   \bibitem[1996]{williams} Williams, R. E., Blacker, B., Dickinson, M., Dixon, W. Van Dyke,  
     Ferguson, H. C., Fruchter, A. S. and 11 coauthors, 1996, 
     \aj, 112, 1335

%%   \bibitem[2005] {wu05} H. Wu, C. Cao, C.N.Hao, F.-S. Liu, J.-L. Wang, X.-Y. Xia, Z.-G. Deng; C. K.-S. Young, 2005, ApJ, 632, L9
  
   \bibitem[2004]{xilouris04} {Xilouris}, E.~M., {Madden}, S.~C., {Galliano}, F., {Vigroux}, L.
     {Sauvage}, M. 2004, A\&A, 416, 41
     
   \end{thebibliography}
\end{document}